\documentclass{article}
\usepackage{arxiv}

\usepackage[utf8]{inputenc}
\usepackage[T1]{fontenc}

\usepackage{url}
\usepackage{graphicx}
\usepackage[hidelinks]{hyperref}

\usepackage{orcidlink}

\usepackage{biblatex}
\bibliography{bibliography}

\usepackage[group-separator={,}]{siunitx}

\usepackage{subcaption}
\DeclareGraphicsExtensions{.pdf}

\usepackage{glossaries}
\makeglossaries
\newacronym{api}{API}{Application Programming Interface}
\newacronym{ceu}{CEU}{Central European University}
\newacronym{cdr}{CDR}{Call Detail Record}
\newacronym{csv}{CSV}{Comma Separated Value}
\newacronym{etl}{ETL}{extract, transform, load}
\newacronym{gis}{GIS}{Geographic Information System}
\newacronym{gps}{GPS}{Global Positioning System}
\newacronym{huf}{HUF}{Hungarian forint} 
\newacronym{imei}{IMEI}{International Mobile Equipment Identity}
\newacronym{ksh}{KSH}{Központi Statisztikai Hivatal, Hungarian Central Statistical Office}
\newacronym{mti}{MTI}{Magyar Távirati Iroda, Hungarian news agency}
\newacronym{osm}{OSM}{OpenStreetMap}
\newacronym{pca}{PCA}{Principal Component Analysis}
\newacronym{ses}{SES}{Social Economic Status}
\newacronym{sim}{SIM}{Subscriber Identity Module}
\newacronym{tac}{TAC}{Type Allocation Code}
\newacronym{umts}{UMTS}{Universal Mobile Telecommunications System}
\newacronym{wgs84}{WGS 84}{World Geodetic System, also known as EPSG:4326}
\newacronym{wkt}{WKT}{Well-known text}

\usepackage[titletoc,title]{appendix}

\begin{document}
\title{Analyzing the Behavior and Financial Status of Soccer Fans from a Mobile Phone Network Perspective: Euro 2016, a Case Study}

\author{Gerg\H{o} Pint\'{e}r \orcidlink{0000-0003-4731-3816} \\
John von Neumann Faculty of Informatics, \\
\'{O}buda University, \\
B\'{e}csi \'{u}t 96/B, 1034 Budapest, Hungary \\
\texttt{pinter.gergo@uni-obuda.hu} \And
Imre Felde \orcidlink{0000-0003-4126-2480} \\
John von Neumann Faculty of Informatics,\\
\'{O}buda University,\\
B\'{e}csi \'{u}t 96/B, 1034 Budapest, Hungary \\
\texttt{felde.imre@uni-obuda.hu}
}

\keywords{mobile phone data; call detail records; type allocation code; data analysis; human mobility; large social event; social sensing; socioeconomic status}

\maketitle{}

\begin{abstract}
In this study, Call Detail Records (CDRs), covering Budapest, for the month of June in 2016 has been analyzed. During this observation period, the 2016 UEFA European Football Championship took place, which affected significantly the habit of the residents, despite the fact that not a single match was played in the city. We evaluated the fans' behavior in Budapest, during and after the Hungarian matches, and found that the mobile phone network activity reflects the football fans' behavior, demonstrating the potential of mobile phone network data within a social sensing system.
The Call Detail Records are enriched with mobile phone properties to analyze the subscribers' devices. Applying the device information (Type Allocation Code) from the activity records, the Subscriber Identity Modules, that do not operate in cell phones are omitted from mobility analyses, allowing to focus on people.
The mobile phone price is proposed and evaluated as a socioeconomic indicator, and correlation between the phone price and the mobility customs have been found. We also found that, beside the cell phone price, the subscriber age and the subscription type also have an effect on the mobility. On the other hand, these do not seem to affect the interest in football.
\end{abstract}

\section{Introduction}
\label{sec:introduction}

Football is one of the most popular sports worldwide and European or World Championships, especially the finals, are among the most watched sporting events. The Euro 2016 Final was watched by more than 20 million people in France \cite{variety2016soccer}, or the
Germany vs. France semifinal was watched by almost 30 million people in Germany \cite{variety2016soccer}. But, what about Hungary?

According to the MTVA (Media Services and Support Trust Fund), that operates the television channel M4~Sport, the first Hungarian match was watched by about 1.734 million people, the second by about 1.976 million and the third group match by about 2.318 million people\footnote{According to the Hungarian Central Statistical Office (\acrshort{ksh}), the population of Hungary was about 9.83 million in 2016 \cite{ksh22.1.1.1}.}. With these ratings, the M4~Sport, turned out to be the most watched television channel in Hungary, during those days \cite{hiradohu2016csoportgyoztes}.
The whole participation of the Hungarian national football team was beyond expectations and raised interest, even among those, who generally, do not follow football matches.
But, is it possible to measure/correlate this interest, with a mobile phone network?

Mobile phones can function as sensors, that detect the whereabouts and movement of their carrier. In this day and age, practically everyone has a mobile phone, that makes it possible to use large scale analyses. With enough data, the general mobility customs and reactions to events can also be studied.
The first step is to prepare the data and select the appropriate individuals for the study.
Filtering the subscribers of the \acrshort{cdr} data sets is always a crucial step. Not just to eliminate the inactive users: a subscriber, who only appears a few times in a data set, cannot be used for mobility analysis, but the abnormally active subscribers can also bias the result. Especially if their location does not change, as \acrshort{cdr} data may not only contain records for cell phones, that are carried by people.

Csáji et al. took into account subscribers who had at least 10 activity during the observation period (15 months) \cite{csaji2013exploring}.
Xu et al. chose to use those subscribers, who had at least one activity record at least half of the days during the observation period \cite{xu2018human}. Pappalardo et al. discarded the subscribers who had only one location, and the individuals have at least half as many calls as hours are in the data set. Furthermore, the abnormally active (more than \num{300} calls per day) \acrshort{sim} cards are excluded \cite{pappalardo2015returners}. In \cite{pinter2021evaluating}, we selected the \acrshort{sim} cards, that have activity at least 20 days (out of 30),
the daily mean activity number is at least 40 on workdays and at least 20 on weekends, but not more than \num{1000}. The upper limit is especially important to remove \acrshort{sim} cards, that possibly operate in mobile broadband modems, for example.
Filtering by activity is not necessarily sufficient to keep only individuals in the data set. Type Allocation Codes (\acrshort{tac}), on the other hand, can determine the type of the device and the exact model of a cell phone.

After the right subscribers have been selected, it is common to determine the home and work locations \cite{vanhoof2018assessing,mamei2019evaluating,pappalardo2021evaluation}, then between these two crucial locations, the commuting trends can be identified.
The commuting is studied between cities \cite{lee2018urban,zagatti2018trip,mamei2019evaluating,barbosa2020uncovering} or within a city \cite{diao2016inferring,jiang2017activity,fiadino2017call,fan2018estimation,ni2018spatial,ghahramani2018mobile,ghahramani2018extracting,pinter2021evaluating}.

Apart from commuting and connectivity analysis, \acrshort{cdr} processing is often used \cite{traag2011social,xavier2012analyzing,mamei2016estimating,furletti2017discovering,marques2018understanding,pinter2019activity,rotman2020using,hiir2020impact} for large social event detection.
When thousands of people are on the same place at the same time, they generate a significant `anomaly' in the data, whereas small groups usually do not stand out from the `noise'. This is especially true when the passive, transparent communication between the mobile phone device and the cell are not included in the data, but only the active communication (voice calls, text messages and data transfer) are recorded.

In \cite{pinter2019activity} and \cite{rotman2020using}, mass protests are analyzed via mobile phone network data.
In \cite{traag2011social,mamei2016estimating,xavier2012analyzing} and \cite{hiir2020impact}, the authors examined the location of stadiums, where the football matches took place. Traag et al. \cite{traag2011social} and Hiir et al. \cite{hiir2020impact} also found that the mobile phone activity of the attendees decreased significantly. In \cite{traag2011social}, z-score is also used to express the activity deviation during the social event from the average. Xavier et al. compared the reported number of attendees of these events with the detected ones. Furletti et al. also analyzed sociopolitical events, football matches and concerts, in Rome \cite{furletti2017discovering}.
This paper focuses on football matches, that however, took place in a remote country (France), and the fans' activity are studied in Budapest.

Mobility indicators, such as Radius of Gyration or Entropy, are often calculated \cite{pappalardo2015returners,xu2018human} to describe and classify the subscribers' mobility customs. Furthermore, using mobility to infer about Social Economic Status (\acrshort{ses}) is a current direction of mobility analysis \cite{xu2018human,cottineau2019mobile,barbosa2020uncovering,pinter2021evaluating}.
Cottineau et al. \cite{cottineau2019mobile} explored the relationship between mobile phone data and traditional socioeconomic information from the national census in French cities.
Barbosa et al. found significant differences in the average travel distance between the low and high income groups in Brazil \cite{barbosa2020uncovering}. Xu et al. \cite{xu2018human} found opposite travel tendencies in mobility of Singapore and Boston. In our previous work \cite{pinter2021evaluating}, we showed that the real estate price of the home and work  locations characterize the mobility and validated our results with census data.
In this paper, the price and the age of the subscribers' mobile phones are proposed as a source of the socioeconomic indicator.
While Blumenstock et al. used the call history as a factor of socioeconomic status \cite{blumenstock2015predicting},
Sultan et al. \cite{sultan2015mobile} applied mobile phone prices as socioeconomic indicator and identified areas where more expensive phones appear more often, however, only manually collected market prices were used.

Mobile phone network data is also used to analyze the human mobility during COVID-19 pandemic and the effectiveness of the restrictions.
Willberg et al. identified a significant decrease of the population presence in the largest cities of Finland after the lockdown compared to a usual week \cite{willberg2021escaping}.
Bushman et al. analyzed the compliance to social distancing in the US using mobile phone data \cite{bushman2020effectiveness}. Gao et al. found negative correlation in stay-at-home distancing and COVID-19 increase rate \cite{gao2020association}.
Still, these analyses might not be common enough. Oliver et al. asked: `Why is the use of mobile phone data not widespread, or a standard, in tackling epidemics?' \cite{oliver2020mobile}. This, however, is not within the scope of this paper.

In this study, we analyzed the mobile phone network activity before, during and after the matches of the Hungarian national football team. The Call Detail Records (\acrshort{cdr}), analyzed in this study, have been recorded Budapest, however the matches took place in France. We present another example of social sensing, using \acrshort{cdr}s, in an indirect and a direct way. Indirectly, as the mobile phone activity of the sport fans, residing in Budapest, are studied during matches played in France. Directly, as the spontaneous festival on the streets of Budapest after the third match, and the welcome event at the Heroes' Square are presented from a data perspective.

The Call Detail Records are filtered by the Type Allocation Codes (\acrshort{tac}) to remove those Subscriber Identity Module (\acrshort{sim}) cards, that do not operate in mobile phones, thus not used by actual people. The price and age of the cell phones are also analyzed in contrast of the subscribers' age and mobility customs.

The contributions of this paper are summarized briefly as follows:
\begin{enumerate}
  \item Fusing \acrshort{cdr} data set with mobile phone prices and release dates.
  \item Filtering out \acrshort{sim} cards, that do not operate in mobile phones.
  \item Demonstrating connection between the phone price and the mobility customs.
  \item Proposing mobile phone price as a \acrshort{ses} indicator.
  \item Attendees of the large social events are compared to the rest of the subscribers based on their mobility and \acrshort{ses}.
\end{enumerate}

The rest of this paper is organized as follows. The utilized data is described in Section~\ref{sec:materials}, then, in Section~\ref{sec:methodology}, the applied methodology is summarized, and in Section~\ref{sec:results}, the results of this study are introduced. Finally, in Section~\ref{sec:conclusions}, the findings of the paper are summarized and concluded.

\section{Materials}
\label{sec:materials}

Vodafone Hungary, one of the three mobile phone operators providing services in Hungary, provided anonymized \acrshort{cdr} data for this study. The observation area was Budapest, capital of Hungary and its agglomeration, and the observation period is one month (June 2016). In 2016 Q2, the nationwide market share of Vodafone Hungary was 25.3\% \cite{nmhh_mobile_market_report}. This data set contains \num{2291246932} records from \num{2063005} unique \acrshort{sim} cards, and does not specify the type of the activity (voice calls, text messages or data transfers).

\begin{figure}[ht]
    \centering
    \includegraphics[width=\linewidth]{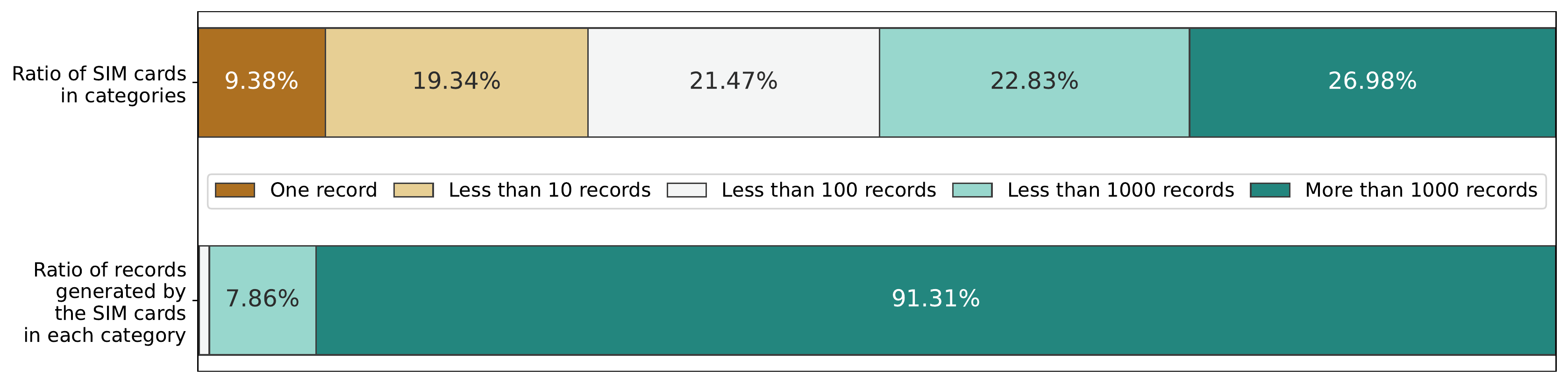}
    \caption{\acrshort{sim} cards categorized by the number of activity records.
    The \acrshort{sim} cards with more than 1000 activity records (26.98\% of the \acrshort{sim} cards) provide the majority (91.31\%) of the activity.}
    \label{fig:vod201606_sim_activity}
\end{figure}

Figure~\ref{fig:vod201606_sim_activity}, shows the activity distribution between the activity categories of the \acrshort{sim} cards. The dominance of the last category, \acrshort{sim} cards with more than 1000 activity records, is even more significant. This almost 27\% of the \acrshort{sim} cards produce the more the 91\% of the activity.

\begin{figure}[ht]
    \centering
    \includegraphics[width=\linewidth]{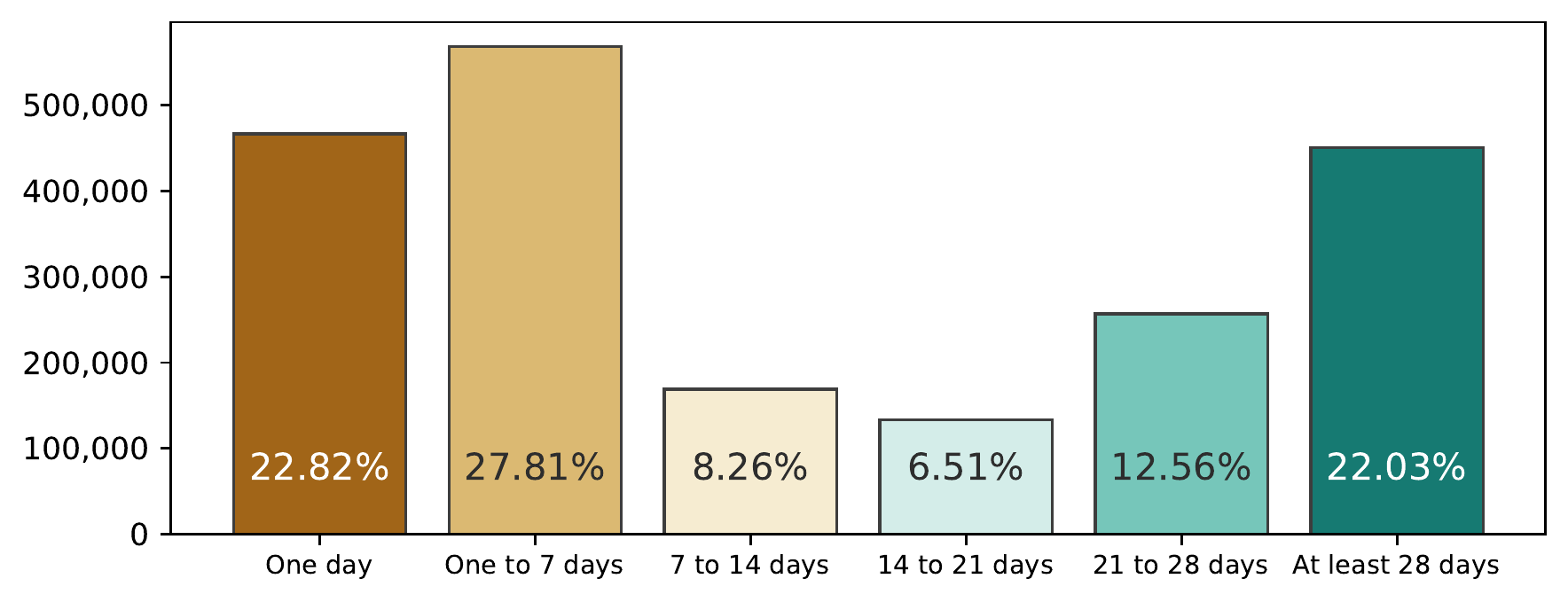}
    \caption{\acrshort{sim} card distribution by the number of active days.}
    \label{fig:vod201606_activity_by_days}
\end{figure}

Figure~\ref{fig:vod201606_activity_by_days}, shows the \acrshort{sim} card distribution by the number of active days. Only the 34.59\% of the \acrshort{sim} cards have activity on at least 21 different days.
There were \num{241824} \acrshort{sim} cards (11.72\%), that have appearance at least two days, but the difference between the first and the last activity is not more the seven days. This may indicate the presence of tourists. High tourism is usual during this part of the year.

The obtained data was in a `wide' format, and contained a
\acrshort{sim} ID, a timestamp, cell ID, the base station (site) coordinates in \acrshort{wgs84} projection, the subscriber (age, sex) and subscription details (consumer/business and prepaid/postpaid) and the Type Allocation Code (\acrshort{tac}) of the device.
The \acrshort{tac} is the first 8 digits of the International Mobile Equipment Identity (\acrshort{imei}) number, allocated by the GSM Association and uniquely identifies the mobile phone model.

\begin{figure}[ht]
    \centering
    \begin{subfigure}[t]{0.49\linewidth}
        \centering
        \includegraphics[width=\linewidth]{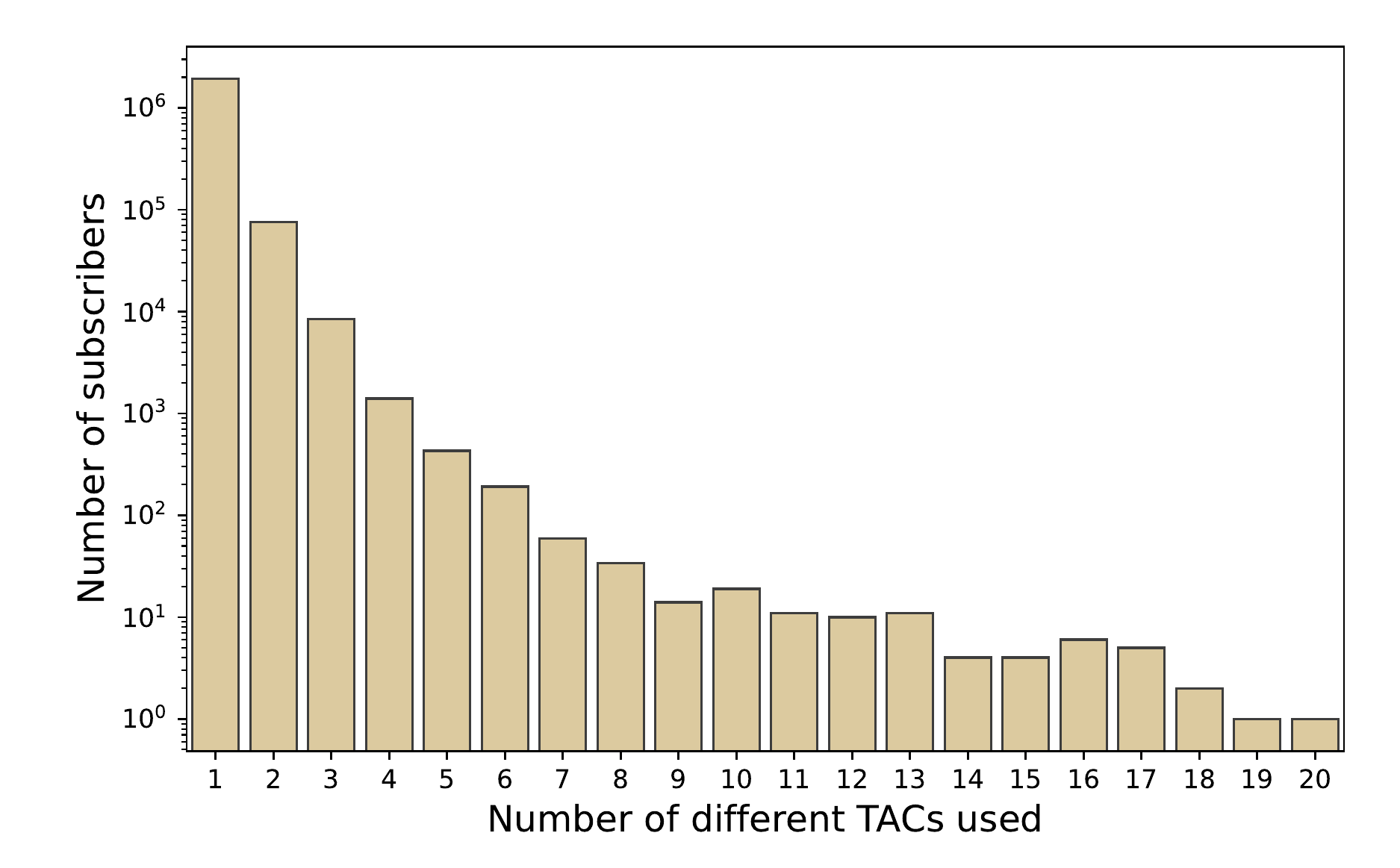}
        \caption{Number of used phones.}
        \label{fig:num_of_diff_tac}
    \end{subfigure}
    \hfill
    \begin{subfigure}[t]{0.49\linewidth}
        \centering
        \includegraphics[width=\linewidth]{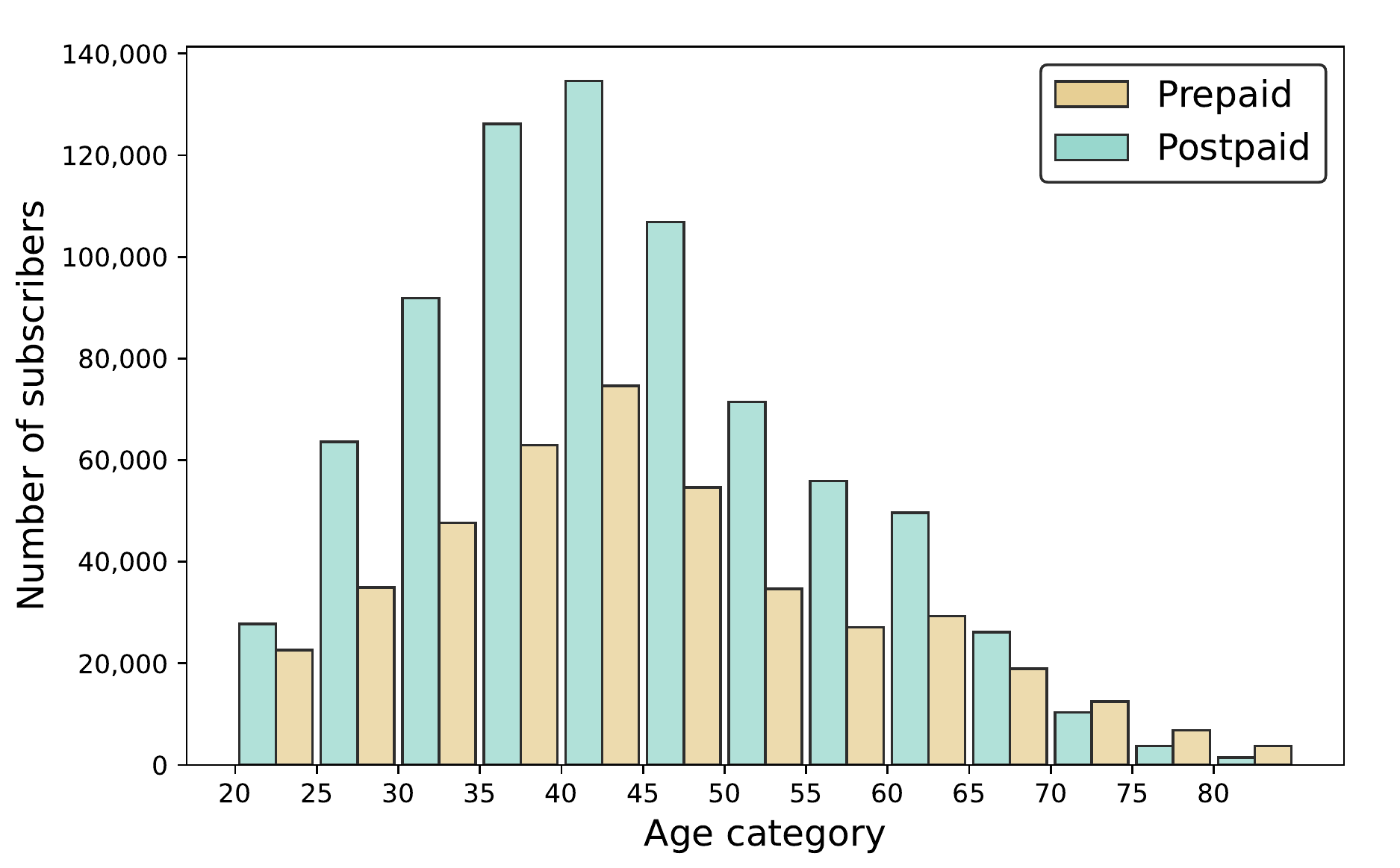}
        \caption{Subscribers' age distribution.}
        \label{fig:age_histogram}
    \end{subfigure}

    \caption{The number of different \acrshort{tac}s used by the subscribers, and the subscriber' age distribution in respect of the subscription type.}
    \label{fig:subscriber_age_device_num}
\end{figure}

The Type Allocation Codes are provided for every record, because a subscriber can change their device at any time. Naturally, most of the subscribers (\num{95.71}\%) use only one device during the whole observation period, but there are some subscribers, maybe mobile phone repair shops, who use multiple devices (see Figure~\ref{fig:num_of_diff_tac}).
As a part of the data cleaning, the wide format has been normalized. The CDR table contains only the \acrshort{sim} ID, the timestamp and the cell ID. A table is formed from the subscriber and the subscription details, and another table to track the device changes of the subscriber.

\begin{figure}[ht]
    \centering
    \includegraphics[width=\linewidth]{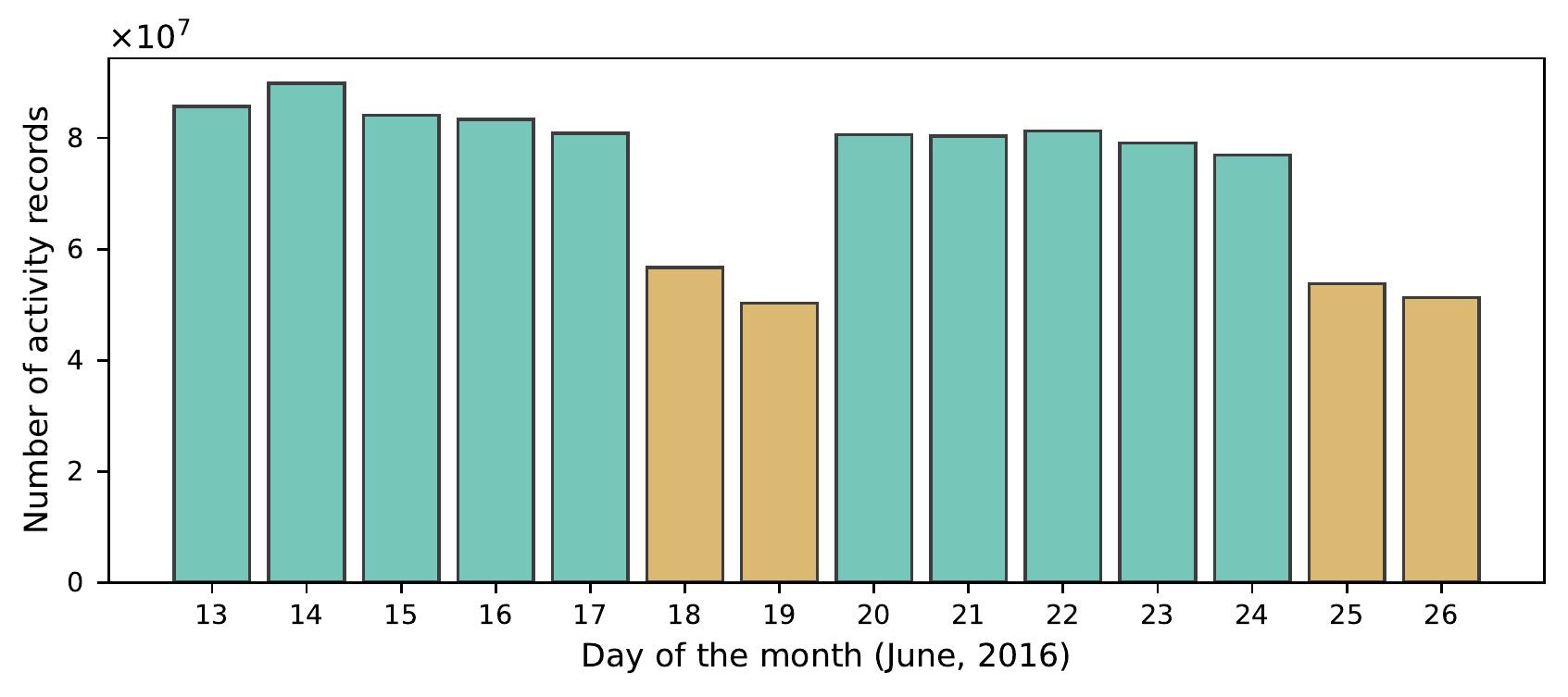}
    \caption{Number of daily activity records, during two weeks of June 2016. The matches of the Hungarian national football team took place from June 14 to June 26.}
    \label{fig:vod201606_daily_activity}
\end{figure}

While the subscription details are available for every \acrshort{sim} cards, the subscriber information is missing in slightly more than 40\% of the cases, presumably because of the subscribers' preferences of personal data usability.
Figure~\ref{fig:age_histogram}, shows the age distribution of the subscribers, whose data is available (\num{58.65}\%), in respect of the subscription type. Note that, this may not represent the age distribution of the population, not even the customers of Vodafone Hungary, as one is allowed to have multiple subscription and the actual user of the phone may differ from the owner of the subscription. Nevertheless, it is still clear that among the elderly people, the prepaid subscriptions are more popular.

Figure~\ref{fig:vod201606_daily_activity}, shows number of daily activity records during the second half of the month. Weekends (brown bars) show significantly fewer activity, hence the activity during the matches compared to the weekday or weekend activity average, respectively to the day of the match.

Although the data contains cell IDs, only the base station locations are known, where the cell antennas are located.
As a base station usually serve multiple cells, these cells has been merged by the serving base stations. After the merge, 665 locations (sites) remained with known geographic locations. To estimate the covered area of these sites, the Voronoi Tessellation, has been performed on the locations. This is a common practice \cite{pappalardo2016analytical,csaji2013exploring,vanhoof2018comparing,candia2008uncovering,novovic2020uncovering,trasarti2015discovering} for \acrshort{cdr} processing.

\subsection{Resolving Type Allocation Codes}

The socioeconomic status \acrshort{ses} of the members in the celebrating crowd have been intended to characterize by the mobile device they use. The preliminary assumption was that the price of the mobile phone represents the \acrshort{ses} of a person.

According to our knowledge, there is no publicly available \acrshort{tac} database to resolve the \acrshort{tac}s to manufacturer and model, although some vendors (e.g., Apple, Nokia) publishes the \acrshort{tac}s of their products.
The exact model of the phone is required to know how recent and expensive a mobile phone is. Although this is not even enough to determine how much the cell phone costed for the subscriber as they could have bought it on sale or discount via the operator in exchange for signing an x-year contract. Still, the consumer price should designate the order of magnitude of the phone price.

The dataset of \acrshort{tac}s provided by ``51Degrees'' has been used, representing the model information with three columns: `HardwareVendor', `HardwareFamily' and `HardwareModel'. The company mostly deals with smartphones that can browse the web, so feature phones and other GSM-capable devices are usually not covered by the data set. Release date and inflated price columns are also included, but these are usually not known, making the data unsuitable to use on its own.

Although it cannot be separated by type, but the \acrshort{cdr} data contains not only call and text message records, but data transfer as well. Furthermore, some \acrshort{sim} cards do not operate in phones, but in other -- often immobile -- devices like a 3G router or a modem. 51Degrees managed to annotate several \acrshort{tac}s as modem or other not phone devices. This was extended by manual search on the most frequent \acrshort{tac}s.
There were \num{324793} \acrshort{sim} cards that uses only one device during the observation period and operates in a non-phone device.

\subsection{Fusing Databases}

For a more extensive mobile phone price database, a scarped GSMArena database\cite{mohit_gsmarena} has been used. GSMArena\footnote{\url{https://www.gsmarena.com/}} has a large and respectable database, that is also used in other studies\cite{reddi2018two,zehtab2021multimodal}. The concatenation of the brand and model fields of the GSMArena database could serve as an identifier for the database fusion. 51Degrees stores the hardware vendor, family and model, where hardware family is often contains a marketing name (e.g., [Apple, iPhone 7, A1778]). As these fields are not always properly distinguished, the concatenation of the three fields may contain duplications (e.g., [Microsoft, Nokia Lumia 820, Lumia 820]).
So, for the 51Degrees records, three identifiers are built using the concatenation of fields (i) vendor + family, (ii) vendor + model and (iii) vendor + family + model, and all the three versions are matched against the GSMArena records.

Another step of the data cleaning is to correct the name changes. For example, BlackBerries were manufactured by RIM (e.g., [RIM, BlackBerry Bold 9700, RCM71UW]), but later, the company name was changed to BlackBerry and the database records are not always consistent in this matter. The same situation occurs due to the Nokia acquisition by Microsoft.

To match these composite identifiers, the simple string equality cannot be used, due to writing distinction, so Fuzzy String match is applied using the FuzzyWuzzy Python package, that uses Levenshtein Distance to calculate the differences between strings. This method is applied for all the three identifiers from the 51Degrees data set and the duplicated matches (e.g., when the family and the model is the same) were removed.
Mapping the GSMArena database to the 51Degrees adds phone price and release date information to the \acrshort{tac}s, that can merged with the \acrshort{cdr}s.

From the GSMArena data, two indicators have been extracted: (i) price of the phone (in EUR), and (ii) the relative age of the phone (in months). The phone price was left intact without taking into consideration the depreciation, and the relative age of the phone is calculated as the difference of the date of the \acrshort{cdr} data set (June 2016) and the release date of the phone.

\section{Methodology}
\label{sec:methodology}

The framework, introduced in our earlier work\cite{pinter2021evaluating}, has been applied to process the mobile phone network data. The \acrshort{cdr}s are normalized, cleaned and the mobility metrics (Section \ref{sec:mobility_metrics}) are determined for every subscriber. The records can be filtered spatially and temporally, both of these filtering is applied for this work. Additionally, a group of \acrshort{sim} cards can be selected from the activity records.

Only temporal filtering is applied to visualize the activity trends during the football matches. Figures \ref{fig:aut_hun_timeseries}, \ref{fig:isl_hun_timeseries}, \ref{fig:hun_prt_timeseries},   \ref{fig:post_match_festival_timeseries}, \ref{fig:hun_prt_activity_fan_activity} and \ref{fig:hun_bel_timeseries}), illustrate the activity of the subscribers in the whole observation area during the matches, including the two hours before and after the matches.
For the celebration after the Hungary vs. Portugal match, spatial and temporal filtering is applied to select the area of interest (Budapest downtown) in the given time interval.

\begin{figure}[t!]
  \centering
  \includegraphics[width=\linewidth]{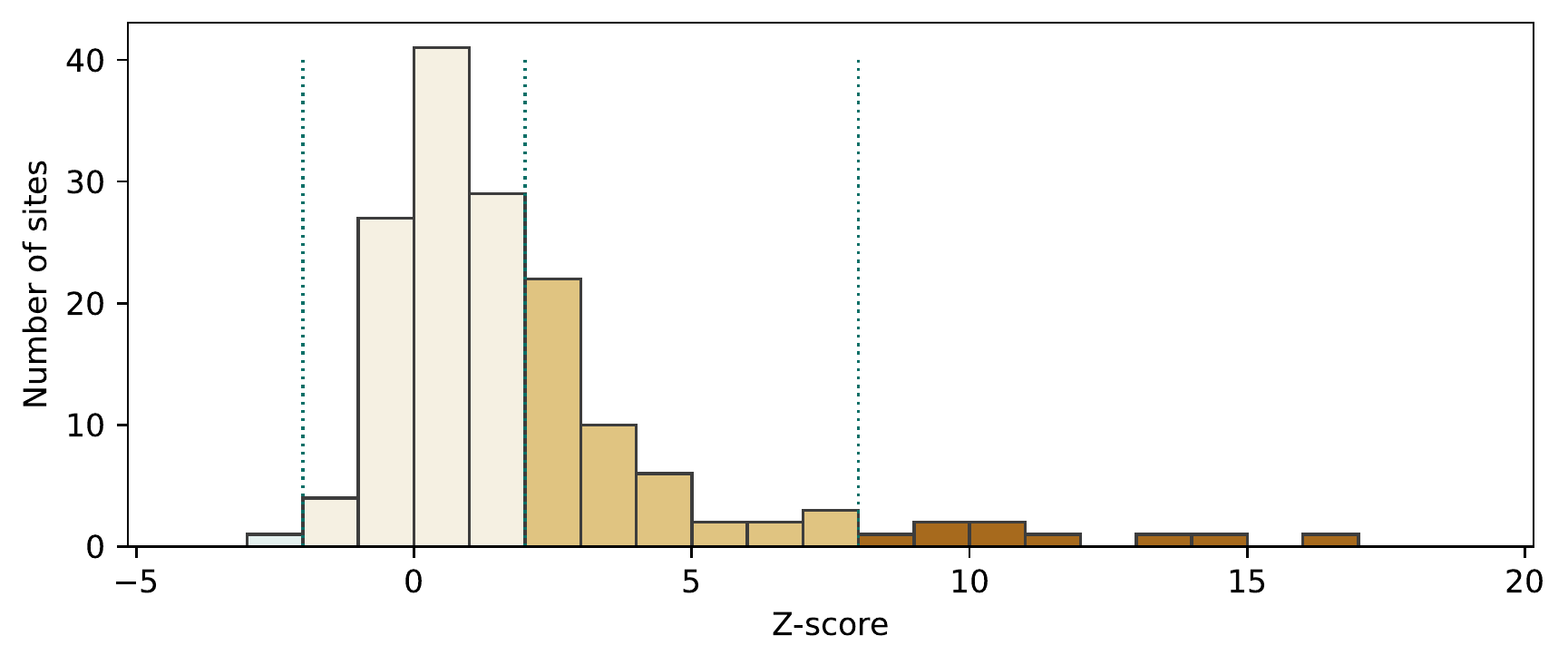}
  \caption{Z-score distribution of the downtown sites, with the activity level thresholds at $-2$, $2$ and $8$, using the same colors as in Figure~\ref{fig:post_match_festival}.}
  \label{fig:zscore_hist}
\end{figure}

To determine the activity levels for the map, Figure~\ref{fig:post_match_festival}, the match-day activity, the average weekdays activity (without the match-day) and the Z-scores
\footnote{The standard score (or z-score) is defined as ${z = \frac{x-\mu}{\sigma}}$, where $\mu$ is the mean and $\sigma$ is the standard deviation.}
are determined for the sites of the area of interest (downtown), in the selected time interval (20:15--20:20). We observed that the standard deviation would be higher, without removing the target-day activity from the reference average, consequently the Z-score would be lower and the relative differences less consistent.
The histogram of the Z-score were generated for the selected sites (Figure~\ref{fig:zscore_hist}) to determine the activity categories. Zero value means that, the activity level equals to the average, but a wider interval (between $-2$ and $2$) is considered average to allow some variation.
Sites with Z-score between $2$ and $8$ are considered having high activity during the given time interval. There are sites with either low (below $-2$) or very high activity (over $8$).
The same method is applied for the map of Figure~\ref{fig:heroes_square_welcoming}, but as the area of interest and the event differs, the thresholds are not the same (see Section~\ref{sec:homecoming}).

The groups of football fans are formed from the subscribers based on only the activity during the Hungary vs. Portugal match. The owner of those non-phone \acrshort{sim} cards, that were active after at least two goals are considered active football fans. The properties of these subscribers, including the age, mobility metrics, phone age and price are compared to the rest of the subscribers (Figure \ref{fig:phone_age_and_price_of_subscribers}).

\subsection{Mobility Metrics}
\label{sec:mobility_metrics}

The metrics of Radius of Gyration and Entropy has been used to characterize human mobility. These indicators are determined for every subscriber, omitting those \acrshort{sim} cards, that operate in non-phone devices. In this study, locations are represented by the base stations.

The Radius of Gyration \cite{gonzalez2008understanding} is the radius of a circle, where an individual (represented by a \acrshort{sim} card) can usually be found.
It was originally defined in Equation~(\ref{eq:gyration}), where $L$ is the set of locations visited by the individual, $r_{cm}$ is the center of mass of these locations, $n_i$ is the number of visits at the i-th location.

\begin{equation}
    \label{eq:gyration}
    r_g = \sqrt{\frac{1}{N} \sum_{i \in L}{n_i (r_i - r_{cm})^2}}
\end{equation}

The entropy characterizes the diversity of the visited locations of an individual's movements, defined as Equation~(\ref{eq:entropy}), where $L$ is the set of locations visited by an individual, $l$ represents a single location, $p(l)$ is the probability of an individual being active at a location $l$ and $N$ is the total number of activities of an individual \cite{pappalardo2016analytical,cottineau2019mobile}.

\begin{equation}
    \label{eq:entropy}
    e = - \frac{\sum_{l \in L}{p(l) \log p}}{\log N}
\end{equation}

\subsection{Socioeconomic Status}
\label{sec:ses}

In our earlier work \cite{pinter2021evaluating}, the real estate price of the subscribers' home locations were used to describe the socioeconomic status.
In this study, the \acrshort{cdr}s are enriched by phone prices and the phone price is assumed to apply as a socioeconomic indicator. To demonstrate the applicability of the mobile phone price as a socioeconomic indicator, it was examined in respect of the mobility indicators, applying Principal Component Analysis (\acrshort{pca}).

The \acrshort{sim} cards are aggregated by the subscriber age categories (5-year steps between 20 and 80) and the phone price categories (100 EUR steps to 700 EUR), the Radius of Gyration and Entropy categories. For the Radius of Gyration, 0.5 km distance ranges are used between 0.5 and 20 km, and the Entropy values are divided into twelve bins between \num{0.05} and \num{1.00}.
The structure of the data used for the Principal Component Analysis defined as follows.

A table has been generated where, every row consists of 40 columns, representing 40 Radius of Gyration bins between 0.5 and 20 km and 20 columns representing 20 Entropy bins, between \num{0.05} and \num{1.00}. The subscribers, belonging to each bin are counted, and the cardinality have been normalized by metrics to be able to compare them.
The categories are not explicitly labeled by them, so the subscriber age and the phone price descriptor columns are not provided to the \acrshort{pca} algorithm.
The same table is constructed using weekend/holiday metrics and its rows are appended after the weekdays ones.

When the \acrshort{pca} is applied, the 60-dimension vector is reduced to two dimensions based on the mobility customs, where the bins are weighted by the number of subscribers. The cumulative variance of the two best components is about 61\% (see Figure~\ref{fig:age_pp_pca_var}). The bins, representing the two new dimensions (PC1 and PC2) are plotted (see Figure~\ref{fig:age_pp_pca}) and the markers are colored by the phone price, marker sizes indicate the subscriber age category, using larger markers for younger subscribers.

\section{Results and Discussion}
\label{sec:results}

\begin{figure}[ht]
    \centering
    \includegraphics[width=\linewidth]{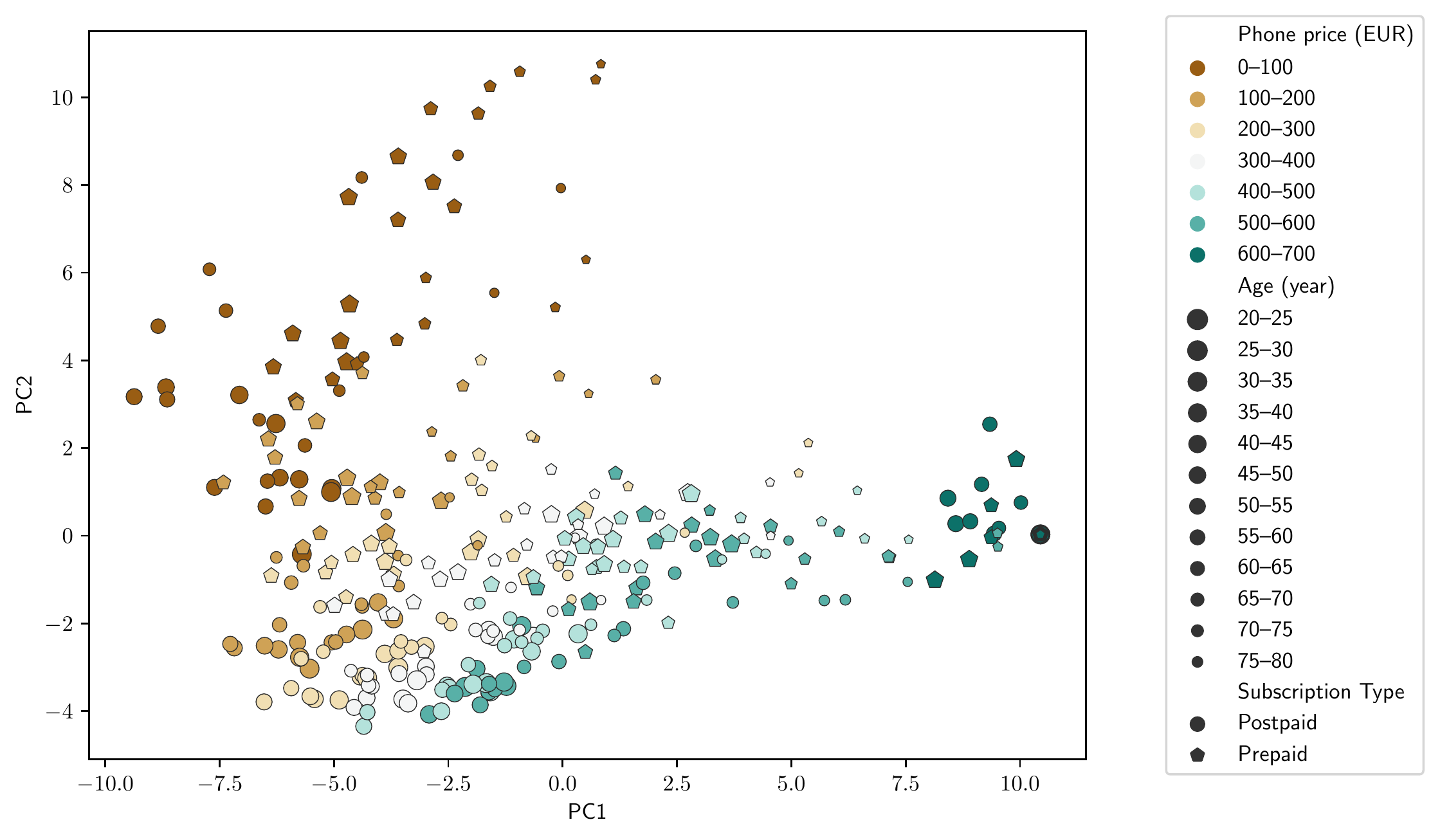}
    \caption{Scatter plot of the 2-component \acrshort{pca}. Marker sizes indicate subscriber age category,
the color represents the phone price category and the subscription type (Prepaid/Postpaid) is distinguished by the marker type.}
    \label{fig:age_pp_pca}
\end{figure}

As Figure~\ref{fig:age_pp_pca} shows, the markers are clustered by color, in other words, the phone price, that is proportional to PC1, but inversely proportional to PC2.
Within each phone price group, the younger subscribers (larger markers) are closer to the origin, indicating that the mobility custom of the younger subscribers differs from the elders, although this difference is smaller within the higher price categories.
This finding coincides with \cite{fernando2018predicting}, where Fernando et al. found correlation between subscribers' age and mobility metrics.

\begin{figure}[ht]
    \centering
    \begin{subfigure}[t]{0.49\linewidth}
        \centering
        \includegraphics[width=\linewidth]{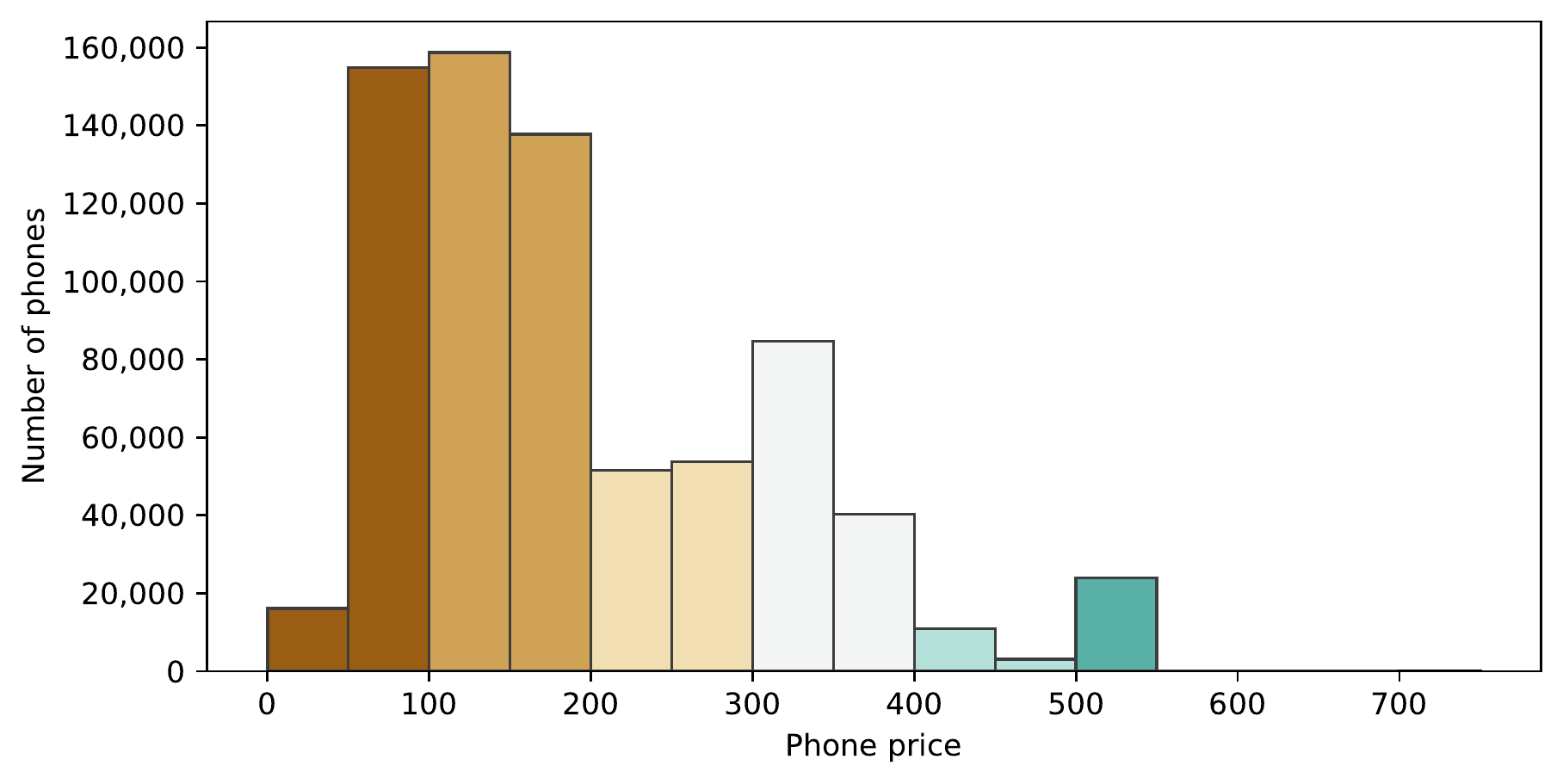}
        \caption{Phone price distribution.}
        \label{fig:pp_hist}
    \end{subfigure}
    \hfill
    \begin{subfigure}[t]{0.49\linewidth}
        \centering
        \includegraphics[width=\linewidth]{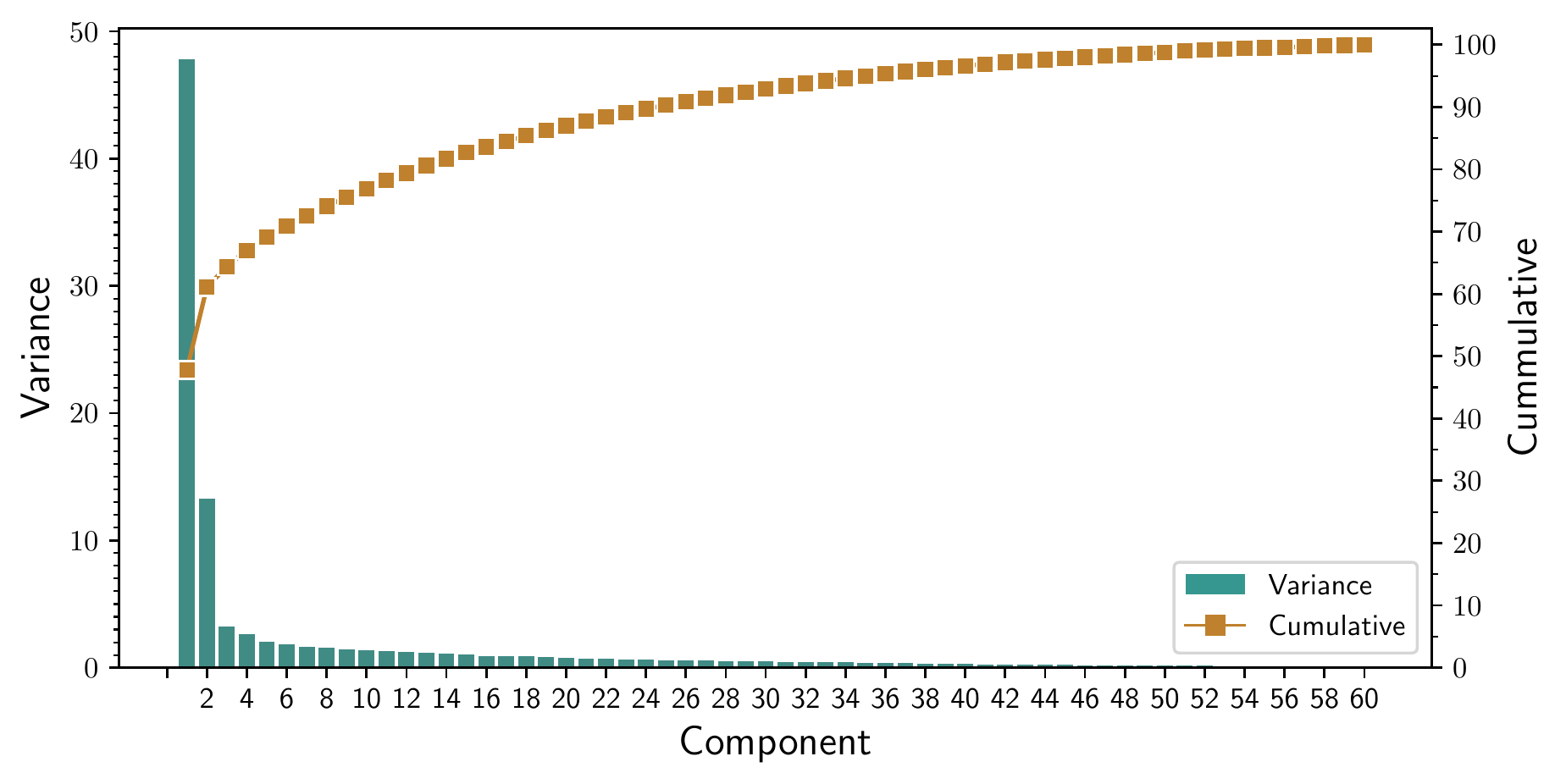}
        \caption{The Pareto histogram for the \acrshort{pca}.}
        \label{fig:age_pp_pca_var}
    \end{subfigure}

    \caption{Phone price distribution and the Pareto histogram for the 60 components of the Principal Component Analysis.}
    \label{fig:pp_dist_and_pca}
\end{figure}

To give context to Figure~\ref{fig:age_pp_pca}, Figure~\ref{fig:pp_hist}, shows the phone price distribution: most of the phones are within the 50--200 EUR range. Note that, there are only a few phones over 550 EUR, but the owners of those have significantly different mobility patterns.

Figure~\ref{fig:age_pp_pca} does not only show that the phone price forms clusters, but also reveals the effect of the subscription type to the mobility. Within the phone price categories, except the highest with only a very few subscribers, the postpaid groups are usually closer to the origin.
Prepaid subscriptions are usually for those, who do not use their mobile phone extensively, and it seems that people with a prepaid subscription have similar mobility customs as people with less expensive phones but postpaid subscription. That is most notable at (-6, 2) and (-5, -1).

\begin{figure}[ht]
    \centering
    \includegraphics[width=.85\linewidth]{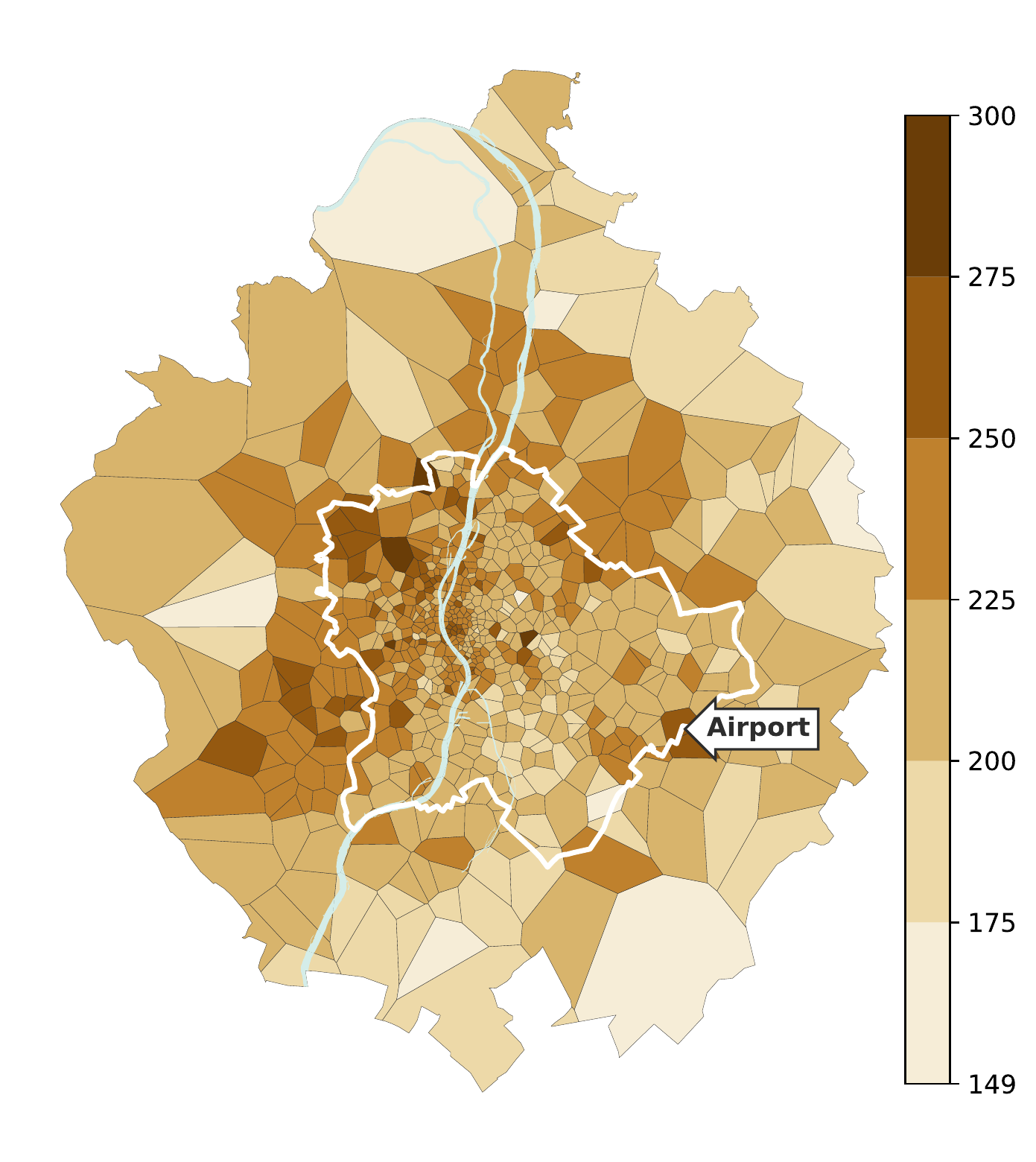}
    \caption{Average price (in EUR) of the mobile phones, that generated the activity records in each site, during the whole observation period (June 2016).}
    \label{fig:avg_phone_price_map}
\end{figure}

Sultan et al. identified areas in Jhelum, Pakistan, where more expensive phones appear more often \cite{sultan2015mobile}. Using the same method, Budapest and its agglomeration was evaluated: the average phone prices from the activity records are determined for every site.
The ground truth is that the real estate prices are higher on Buda side (West of river Danube) of Budapest and downtown \cite{pinter2021evaluating}, and this tendency can be clearly seen in Figure~\ref{fig:avg_phone_price_map}. The airport area has a significantly higher average than its surroundings, that is not surprising.
The spatial tendencies of the mobile phone price, along with the result of the \acrshort{pca} (Figure~\ref{fig:age_pp_pca}), clearly demonstrates the expressiveness of the phone price as a socioeconomic indicator.

The rest of this section examines the results, in the time order of the Hungarian Euro 2016 matches.

\subsection{Austria vs. Hungary}

\begin{figure}[ht]
    \centering
    \includegraphics[width=\linewidth]{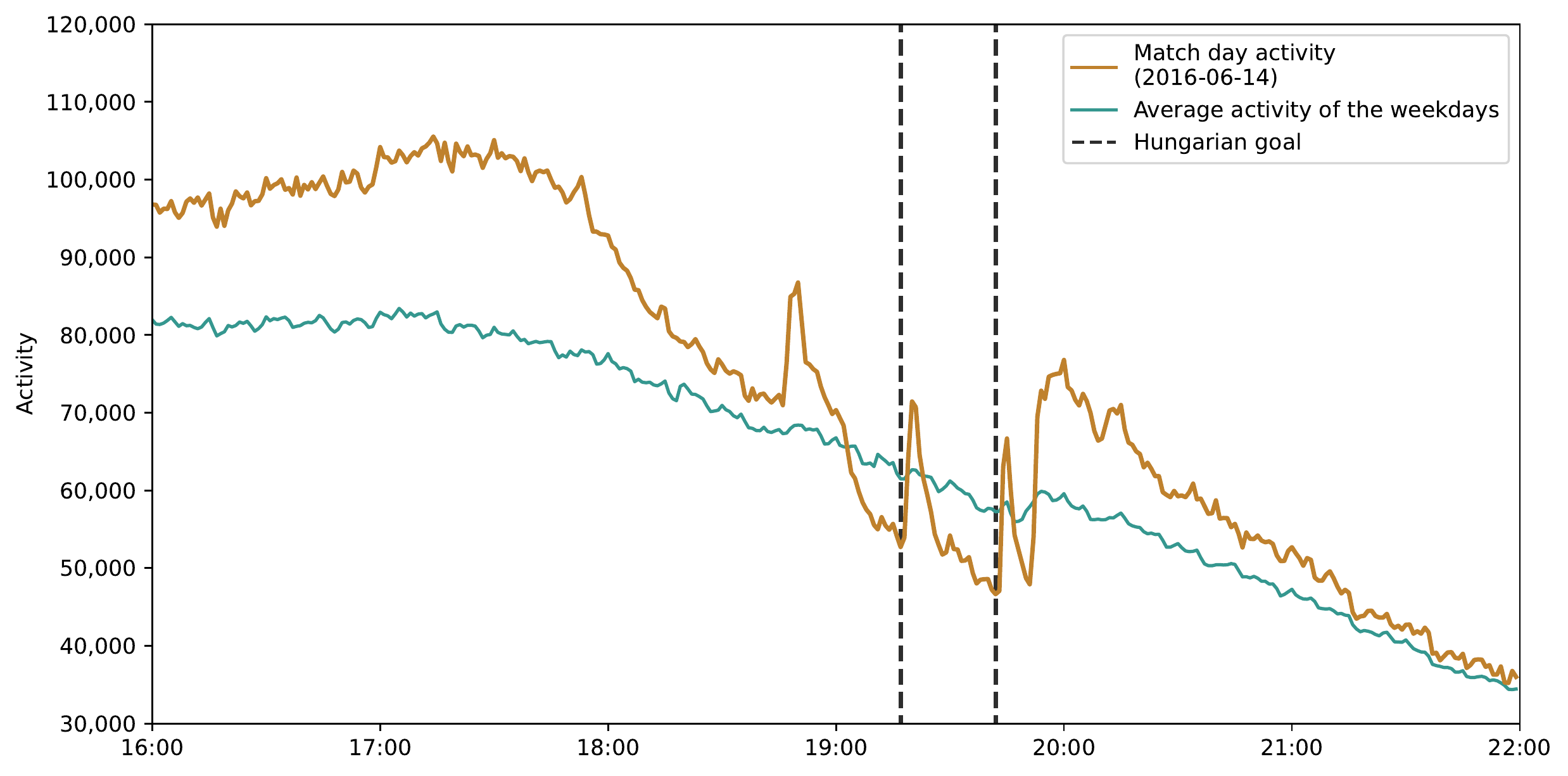}
    \caption{Mobile phone activity during and after the Austria--Hungary Euro 2016 match, in comparison with the average activity of the weekdays.}
    \label{fig:aut_hun_timeseries}
\end{figure}

The first match against Austria (Figure~\ref{fig:aut_hun_timeseries}) was started at 18:00, on Tuesday, June 14, 2016. Before the match, the activity level was significantly higher than the average of the weekdays, and later decreased until the half-time. During the second half, the activity level dropped to the average, which indicated that more people started to follow the match. Right after the Hungarian goals, there are two significant peaks have been observed in the activity, which exactly indicates increased attention and the massive usage of mobile devices during the match.

As the data source cannot distinguish the mobile phone activities by type, it cannot be examined what kind of activities caused the peaks. It is supposed that the activity was mostly data transfer or text messages, not phone calls. It simply does not seem to be lifelike to call someone during the match just because of a goal, but sending a line of text via one of the popular instant messaging services, is very feasible.

\subsection{Iceland vs. Hungary}

\begin{figure}[ht]
    \centering
    \includegraphics[width=\linewidth]{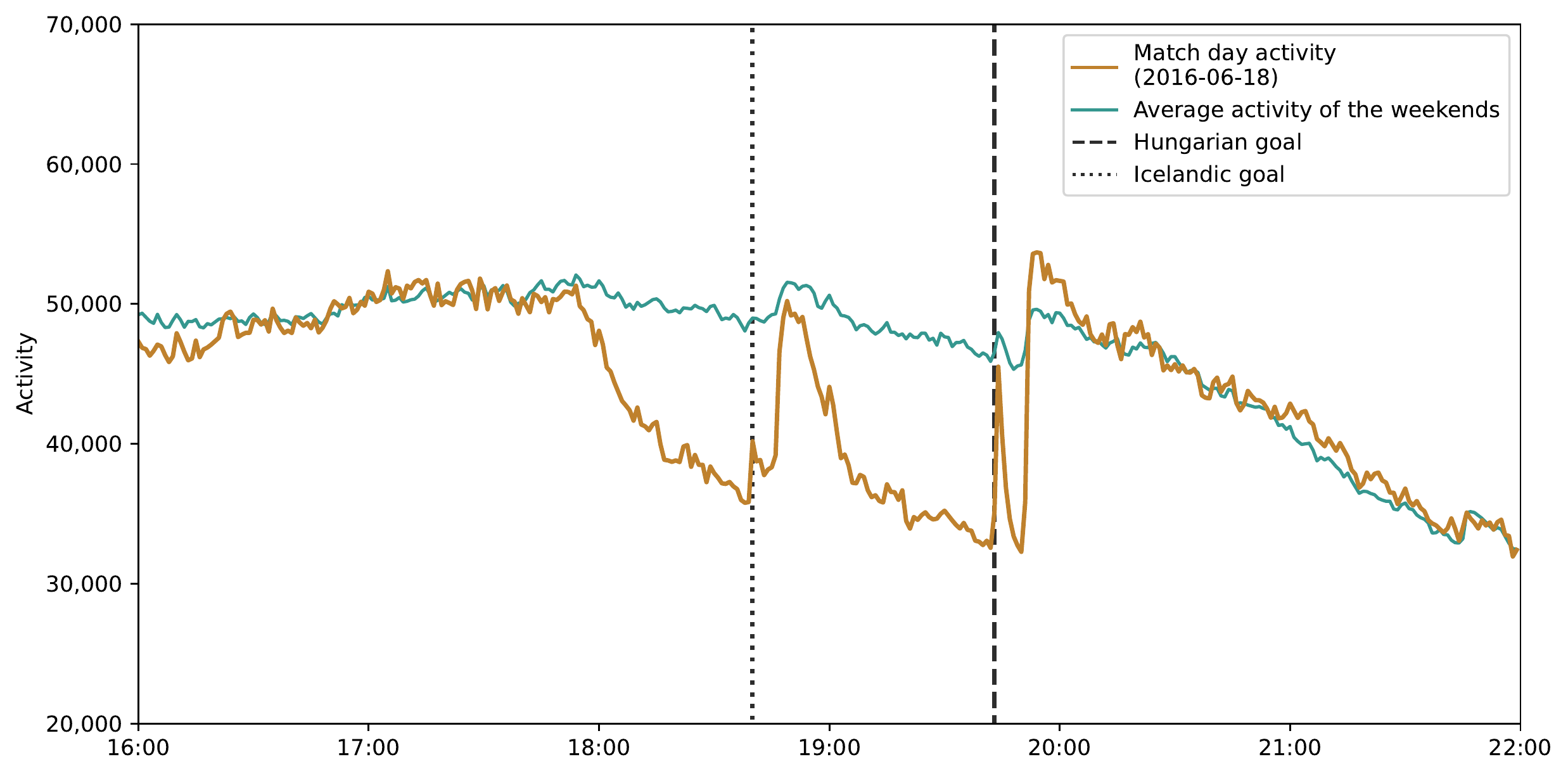}
    \caption{Mobile phone activity during and after the Iceland--Hungary Euro 2016 match, in comparison with the average activity of the weekends.}
    \label{fig:isl_hun_timeseries}
\end{figure}

The match against Iceland was played on Saturday, June 18, 2016. Figure~\ref{fig:isl_hun_timeseries}, shows the mobile phone activity levels before, during and after the match. As the weekend activity is generally lower (see Figure~\ref{fig:vod201606_daily_activity}), the average of the weekdays are used as a reference. The match began at 18:00, and from that point, the activity level was significantly below the average, except the half-time break and, again, the peak after the Hungarian goal.
Interestingly, the Icelandic goal does not result such a significant peak, only a very moderate one can be seen in the time series.

Traag et al. \cite{traag2011social} also found activity drop during a game, but in that case the area of the stadium was analyzed, where the match was played and there was no peak during the match.

\subsection{Hungary vs. Portugal}

\begin{figure}[ht]
    \centering
    \includegraphics[width=\linewidth]{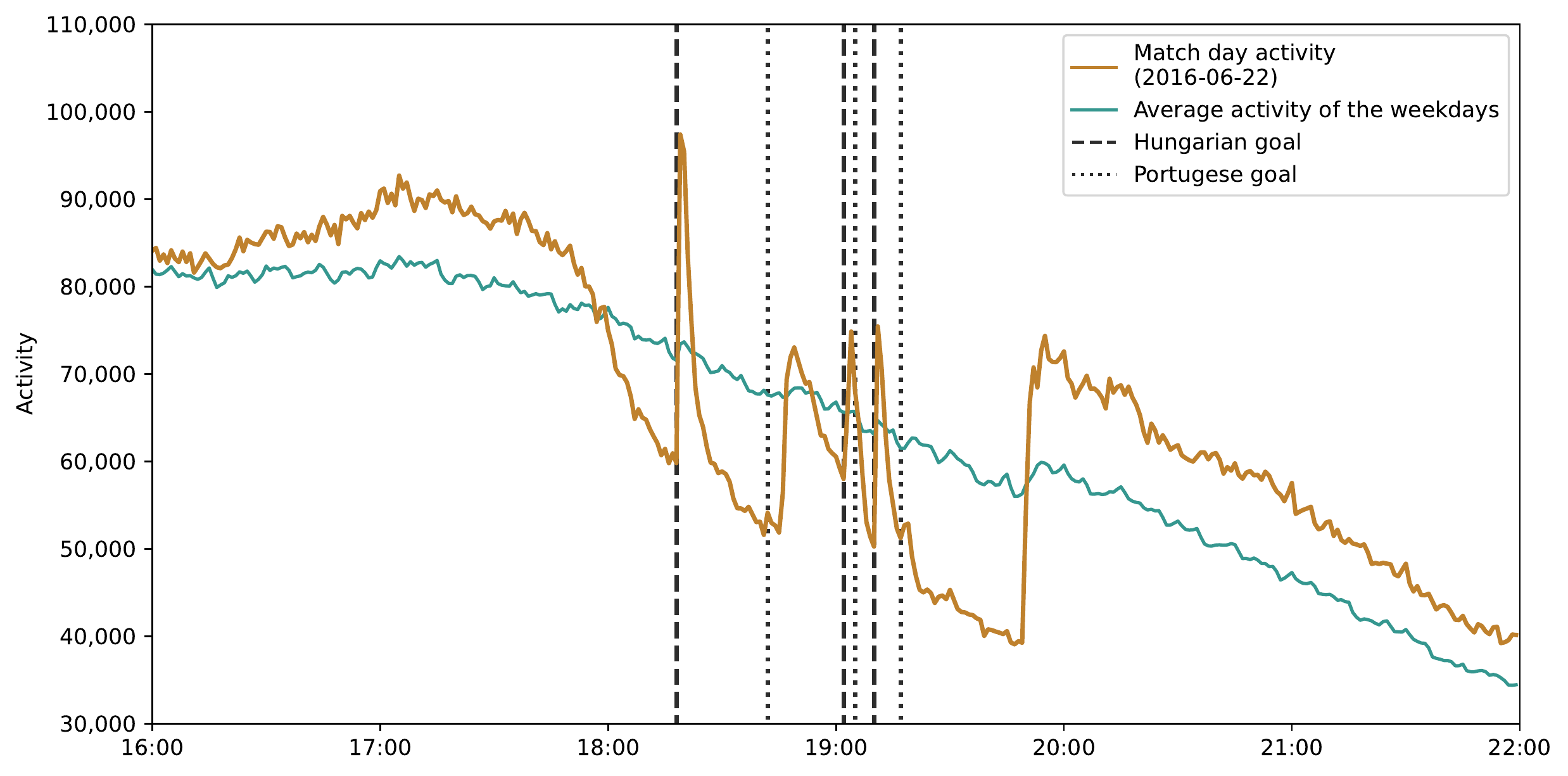}
    \caption{Mobile phone activity during and after the Hungary--Portugal Euro 2016 match, in comparison with the average activity of the weekdays.}
    \label{fig:hun_prt_timeseries}
\end{figure}

On Wednesday, June 22, 2016, as the third match of the group state of the 2016 UEFA European Football Championship, Hungary played draw with Portugal.
Both teams scored three goals and with this result, Hungary won their group and qualified for the knockout phase.
During the match, the mobile phone activity dropped below the average, but the goals against Portugal resulted significant peaks, especially the first one (see Figure~\ref{fig:hun_prt_timeseries}). On the other hand, the Portuguese equalizer goal(s) did not cause significant mark in the activity. In the second half, the teams scored four goals in a relatively short time period, but only the Hungarian goals resulted in peaks.
This observation suggests that the football fans had notable influence on the mobile network traffic.

\begin{figure}[ht]
    \centering
    \includegraphics[width=\linewidth]{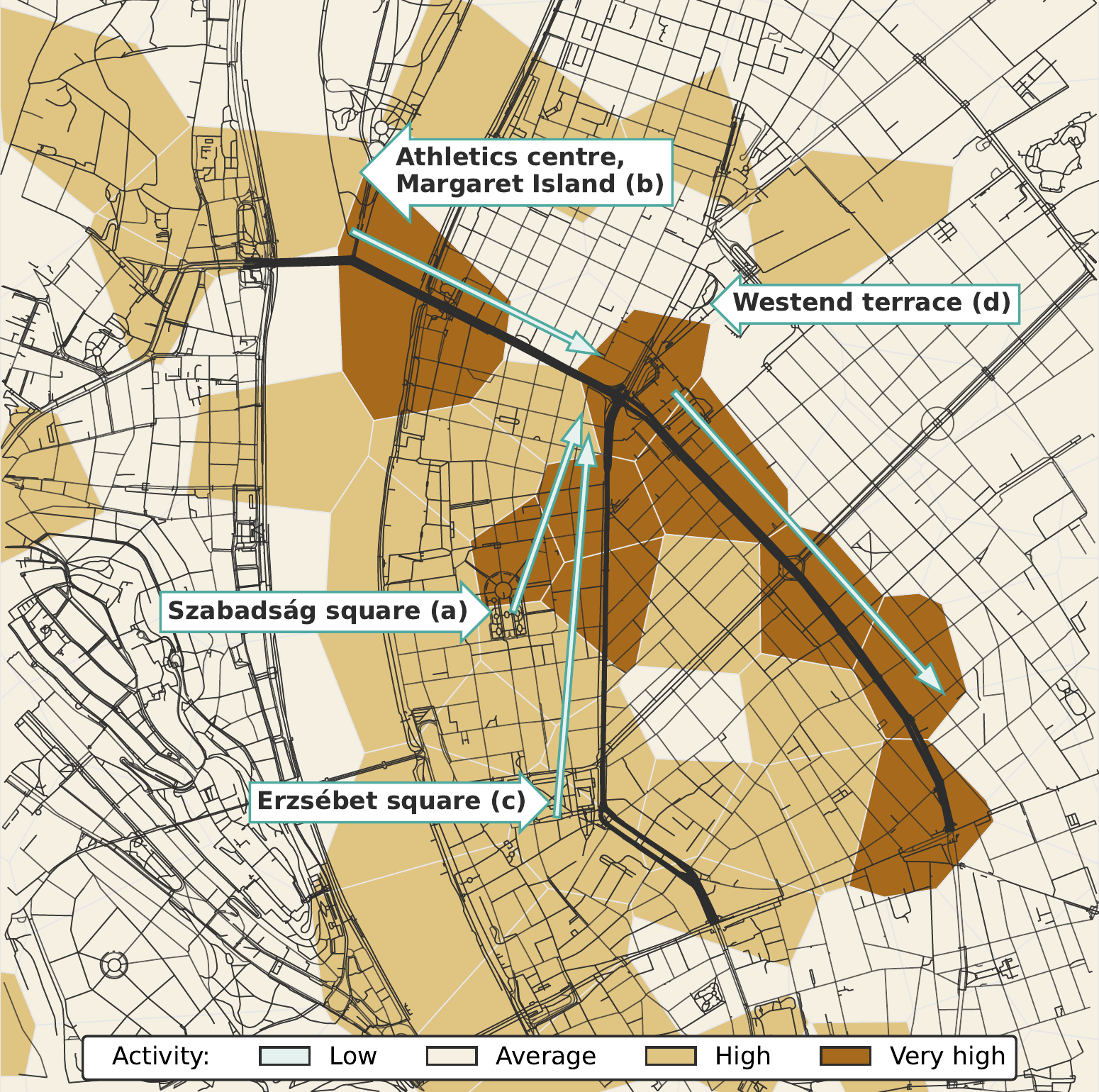}
    \caption{After the Hungary vs. Portugal football match, the fans, delirious with joy, filled the streets. The arrows show their route from the main fan zones to and along the Grand Boulevard. Voronoi polygons colored by the mobile phone network activity at the peak of the event, at 20:15.}
    \label{fig:post_match_festival}
\end{figure}

After the match, the activity level is over the average, that might represent the spontaneous festival in downtown Budapest. According to the \acrshort{mti} (Hungarian news agency), thousands of people celebrated in the streets, starting from the fan zones, mainly from Erzsébet square (Figure~\ref{fig:post_match_festival} a), Margaret Island (Figure~\ref{fig:post_match_festival} b) and Erzsébet square (Figure~\ref{fig:post_match_festival} c) direction Budapest Nyugati railway station. The Grand Boulevard was completely occupied by the celebrating crowd and the public transportation was disrupted along those affected lines.

This social event is comparable to mass protests from a mobile phone network perspective. In an earlier work \cite{pinter2018analysis}, we have analyzed the mobile phone activity at the route of a mass protest. The activity of the cells were significantly high when the protesters passed through the cell. In this case, however, the affected area were smaller and the sites along the Grand Boulevard were very busy at the same time after the game.

\begin{figure}[ht]
    \centering
    \includegraphics[width=\linewidth]{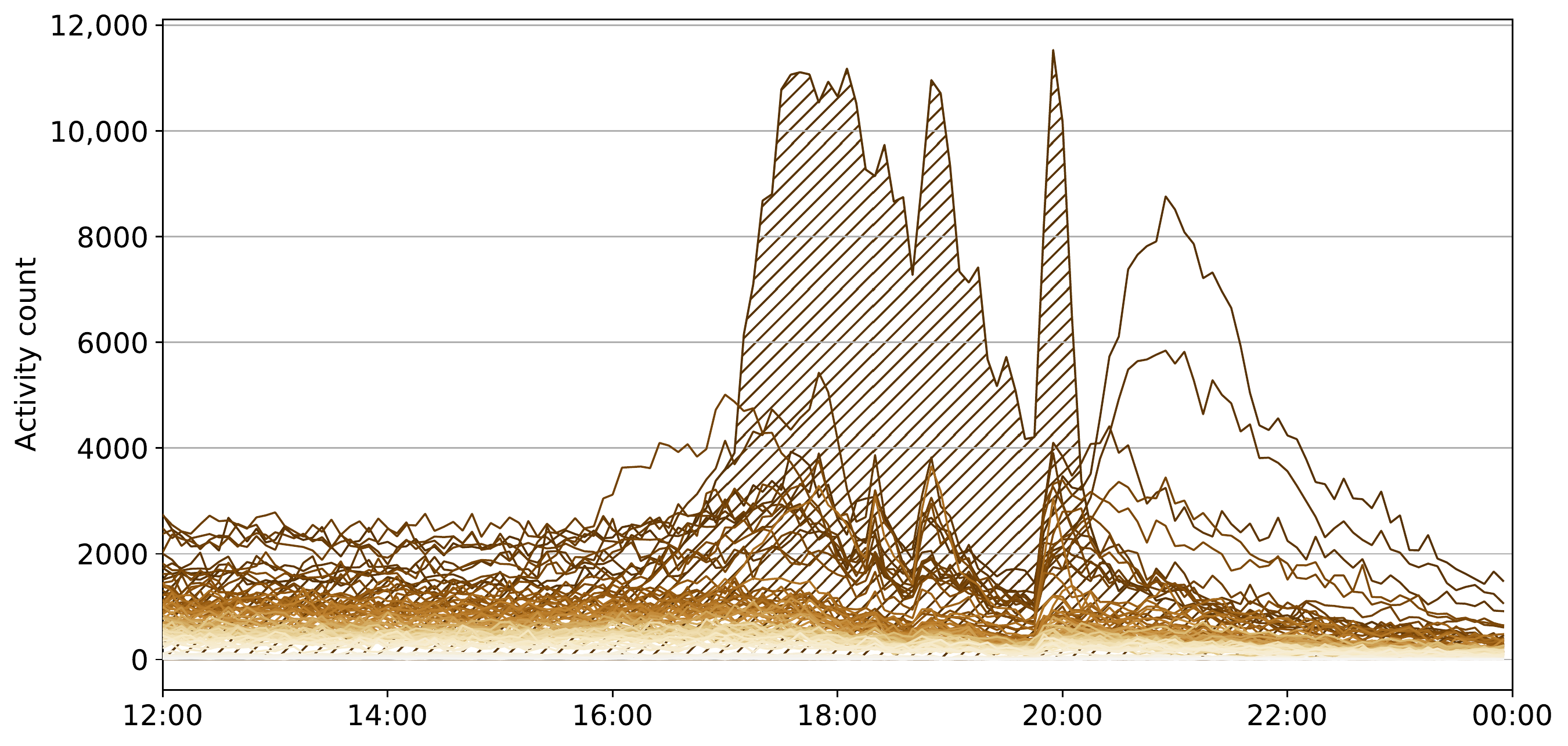}
    \caption{Site activities, in Budapest downtown, on the day of the Hungary vs. Portugal football match (June 22, 2016). The highlighted site covers mostly the Szabadság Square, where one of the main fan zones was set up.}
    \label{fig:post_match_festival_timeseries}
\end{figure}

The activities of the sites (multiple cells aggregated by the base stations), in Budapest downtown, are illustrated on Figure~\ref{fig:post_match_festival_timeseries}.
The highlighted site covers mostly Szabadság square (for the location, see Figure~\ref{fig:post_match_festival} a), where one of the main fan zones was set up with a big screen. The activity curve actually follows the trends of the whole data set (see Figure~\ref{fig:hun_prt_timeseries}). There is high activity before the match, during half-time and, for a short period, after the match. During the match, the activity decreased except four, not so significant, peaks around the goals.

In the highlighted site, in Figure~\ref{fig:post_match_festival_timeseries}, almost 7 thousand \acrshort{sim} cards had been detected between 17:00 and 20:00. The data shows that 53.57\% of the subscribers were between 20 and 50 years old, while 33.49\%, had no age data.

After the match, there is a significant increase in the activity in some other sites. These sites are (mostly) around the Grand Boulevard, where the fans marched and celebrated the advancement of the national football team to the knockout phase.

Figure~\ref{fig:post_match_festival}, shows the spatial distribution of this social event, using Voronoi polygons generated around the base stations locations.
The polygons are colored by the mobile phone network activity increase at 20:15, compared to average of the weekday activity. For the comparison, the standard score
was determined for every base station with a 5-minute temporal aggregation. The darker colors indicate the higher activity surplus in an area.
The figure also denotes the three main fan zones in the area, routes of the fans by arrows, and the affected streets are emphasized.

\subsubsection*{Who are responsible for the peaks?}

There were three Hungarian goals during the match, hence there were three peaks, starting at 18:18, 19:02 and 19:18. All of them had about 5-minute fall-times. To answer this question, the \acrshort{sim} cards that were active during any two of the peaks were selected. Selecting \acrshort{sim} cards that were active during any of the peaks, would also include many subscribers, that cannot be considered as football fan. The participation of all the three peaks, on the other hand, would be too restrictive.

\begin{figure}[ht]
    \centering
    \begin{subfigure}[t]{0.49\linewidth}
        \centering
        \includegraphics[width=\linewidth]{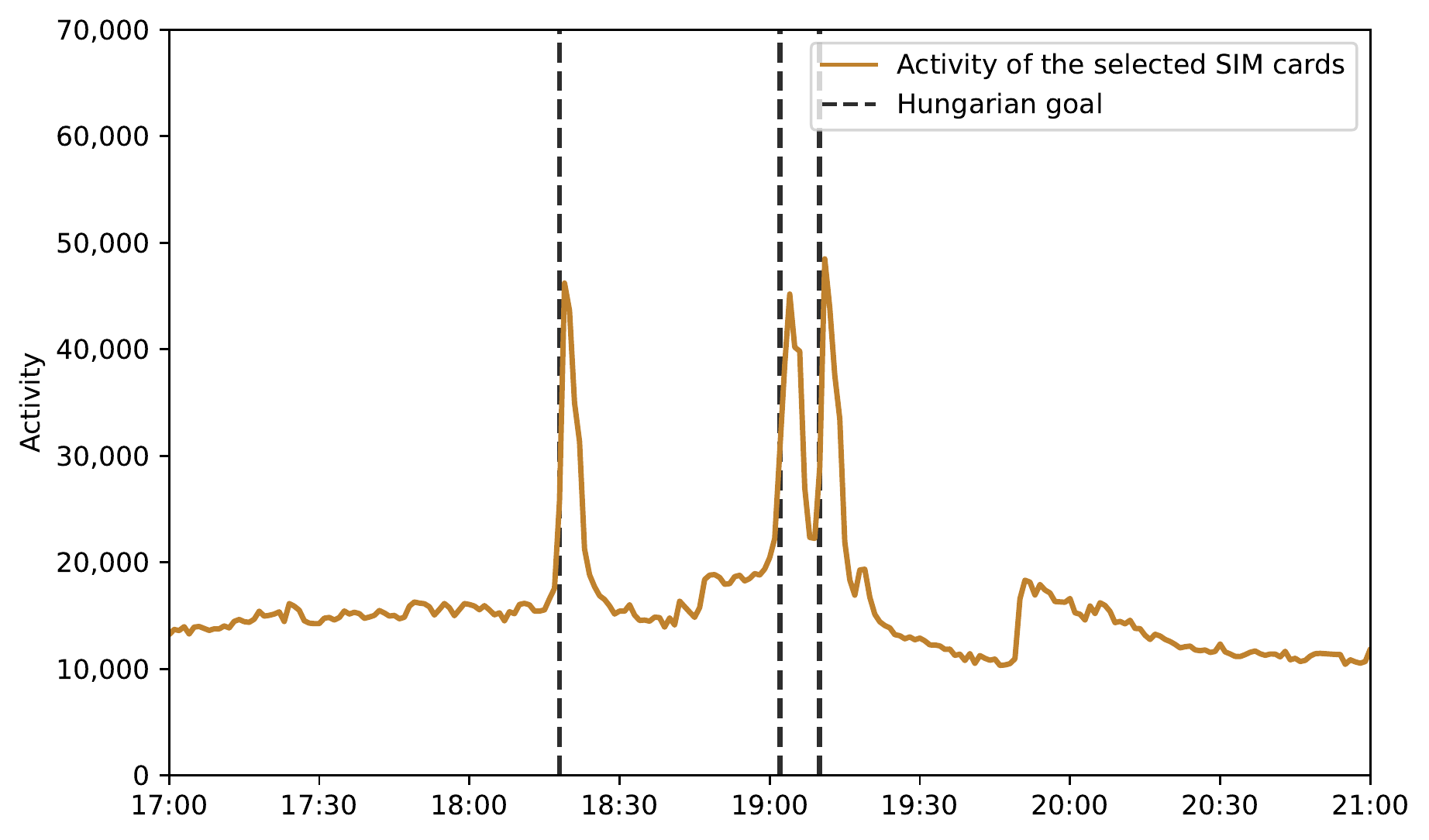}
        \caption{Activity of fans.}
        \label{fig:hun_prt_activity_of_fans}
    \end{subfigure}
    \hfill
    \begin{subfigure}[t]{0.49\linewidth}
        \centering
        \includegraphics[width=\linewidth]{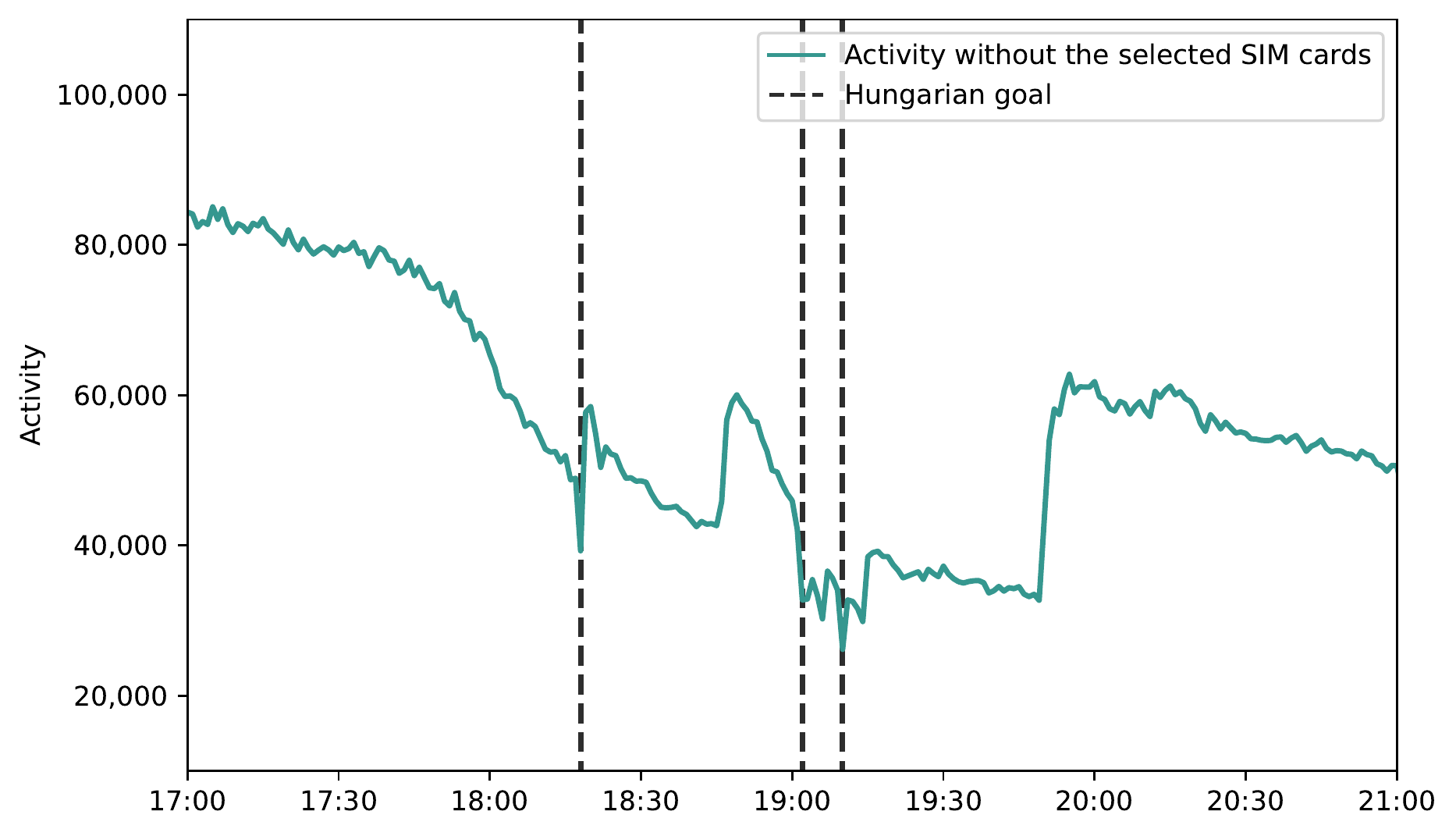}
        \caption{Activity without the fans.}
        \label{fig:hun_prt_activity_without_fans}
    \end{subfigure}

    \caption{Mobile phone network activity of the \acrshort{sim} cards (fans), that had activity right after any two of the Hungarian goals, and the mobile phone activity of the other \acrshort{sim} cards.}
    \label{fig:hun_prt_activity_fan_activity}
\end{figure}

Figure~\ref{fig:hun_prt_activity_of_fans}, presents the activity of the selected \num{44646} \acrshort{sim} cards and the owner of these cards, which may belong to the football fans.
Removing these \acrshort{sim} cards from the data set, should result an activity curve without peaks, and at the same time similar, in tendency, to the average activity. However, as Figure~\ref{fig:hun_prt_activity_without_fans} shows, the activity still drops during the match. Therefore, the `football fan' category should be divided to `active' and `passive' fans, from the mobile phone network perspective. Active fans are assumed to express their joy using the mobile phone network (presumably to access the social media) and cause the peaks. It seems that the passive fans ceased the other activities and watched the game, that caused some lack of activity, compared to the average.
By removing the active fans from the observed set of \acrshort{sim} cards, the activity level decreased in general (Figure~\ref{fig:hun_prt_activity_without_fans}). However, this is not surprising, as these people reacted to the goals, they must often use the mobile phone network. There are also some negative peaks, indicating that the selection is not perfect.

\begin{figure}[ht]
    \centering
    \begin{subfigure}[t]{0.49\linewidth}
        \centering
        \includegraphics[width=\linewidth]{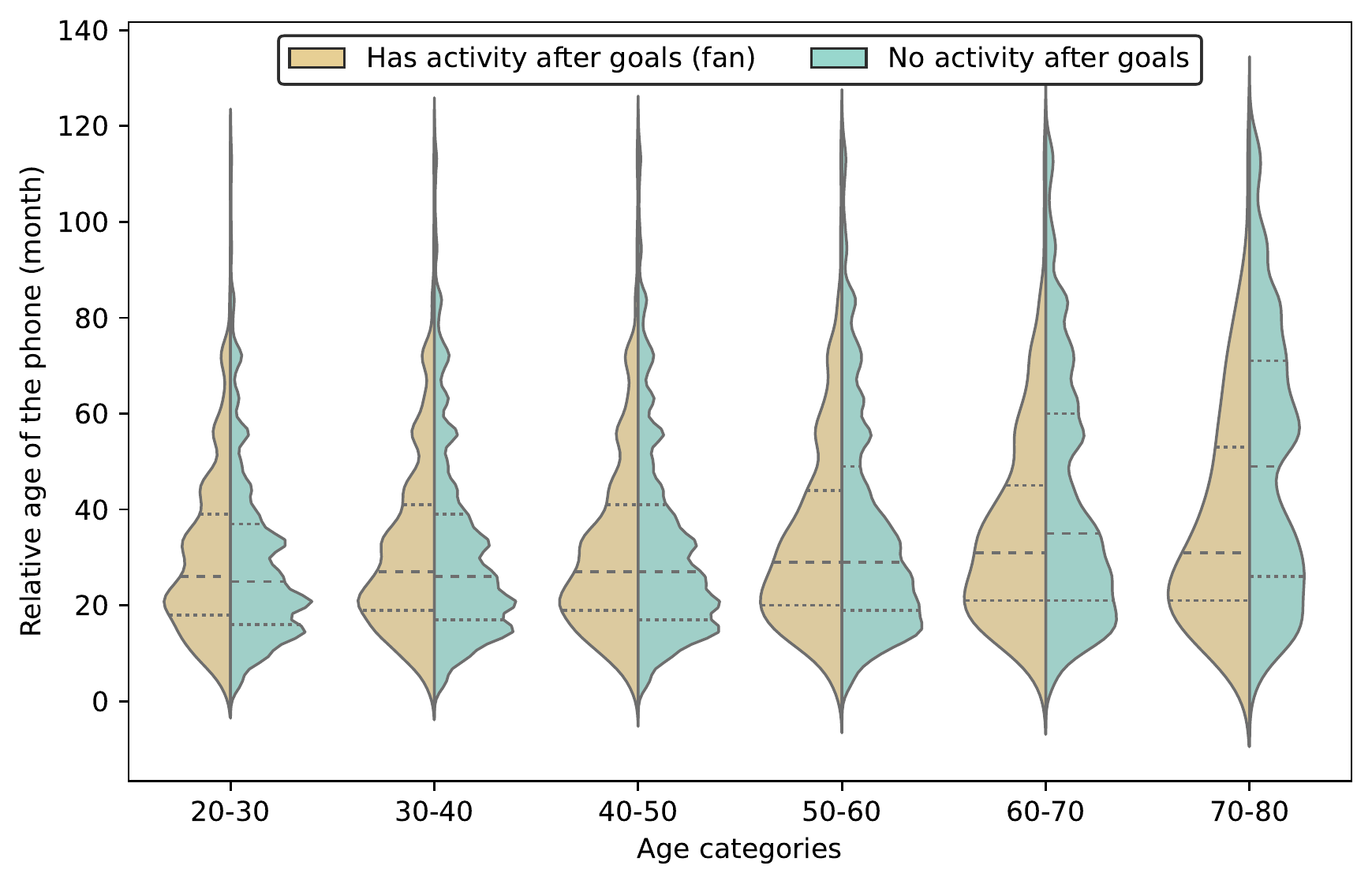}
        \caption{Relative age of the phones.}
        \label{fig:phone_age_of_subscribers}
    \end{subfigure}
    \hfill
    \begin{subfigure}[t]{0.49\linewidth}
        \centering
        \includegraphics[width=\linewidth]{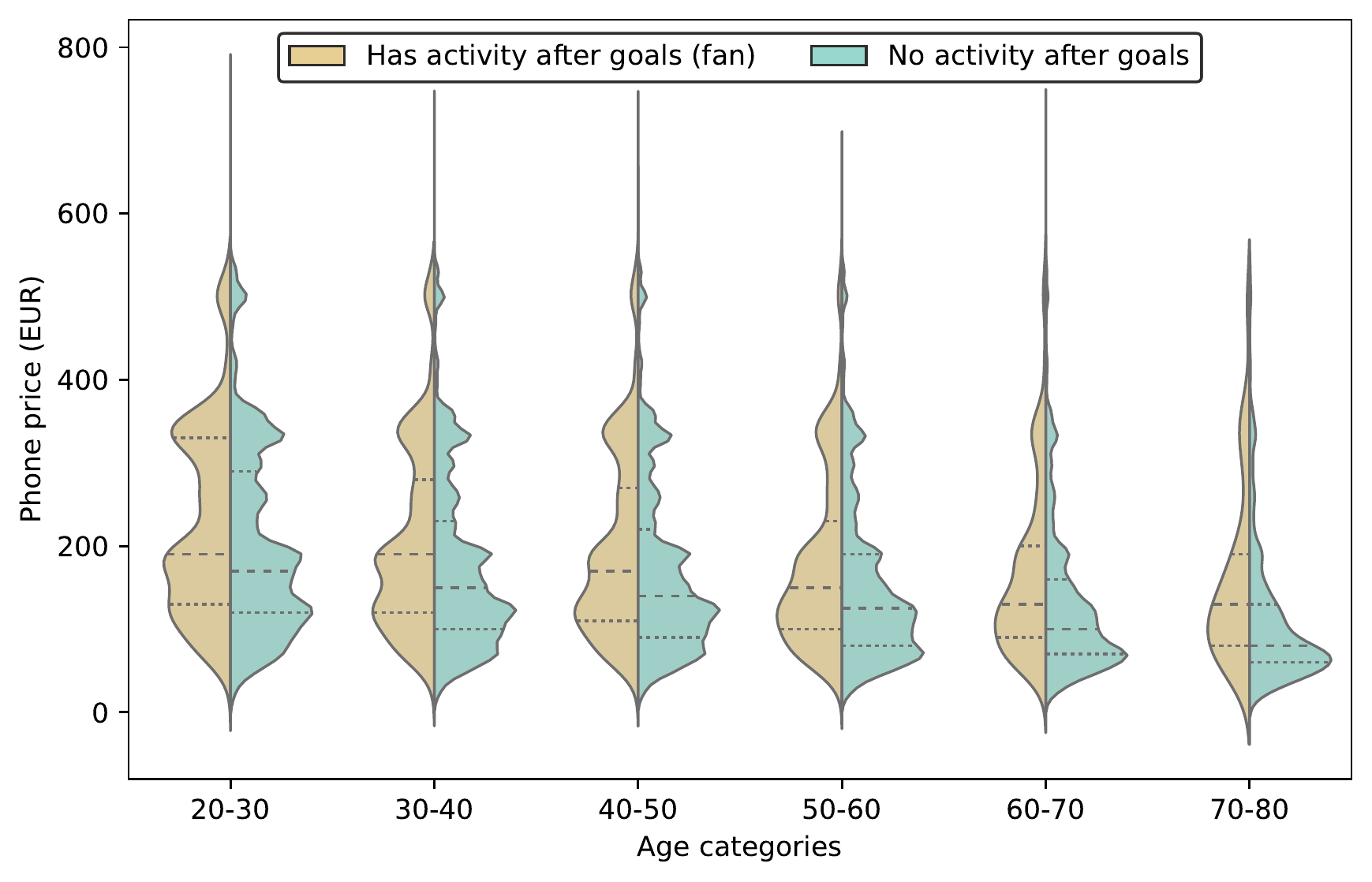}
        \caption{Price of the Phones.}
        \label{fig:phone_price_of_subscribers}
    \end{subfigure}

    \caption{Mobile phone relative age and the price distributions in different age categories, comparing the fans, who had activity right after any two of the Hungarian goals, and the rest of the \acrshort{sim} cards.}
    \label{fig:phone_age_and_price_of_subscribers}
\end{figure}

Is there any difference between the active fans regarding the phone age and price compared to the other subscribers? Figure~\ref{fig:phone_age_of_subscribers}, shows the relative age of the phones in respect of the subscribers' behavior after the goals. No significant difference has been realized between the active fans and other subscribers, the median of the phone relative age is about two years, and there are some much older (nearly ten years old) phones in use. It should be noted that older devices are used by elderly people. The price of the phones show opposite tendency: the younger subscribers own more expensive phones (Figure~\ref{fig:phone_price_of_subscribers}).

Naturally, not all of these \num{169089} \acrshort{sim} card (without the ones operating non-phone devices) generated activity after all the goals. \num{83352} devices were active after the first goal, \num{70603} after the second and \num{68882} after the third. After at least two goals \num{44646}, and after all the three goals only \num{9102} devices had activity, within 5 minutes.

Why would they use the mobile phone network to access social media? If they were at home, they would have used the wired connection, via Wi-Fi for mobile devices. In Hungary, the \num{79.2}\% of the households had wired internet connection, according to the \acrshort{ksh}\cite{ksh12.8.1.9}, and it could be even higher in Budapest. However, if they were at fan zones, for example in Szabadság Square, using the mobile network is more obvious.

As Figure~\ref{fig:phone_age_and_price_of_subscribers} shows, there is no significant difference in the phone age between the active football fans and the rest of the subscribers. The medians are almost the same within the young adult and the middle-age categories, but elders tend to use older devices, especially those, who did not react to the goals.
The active football fans' median phone price is 180 EUR, in contrast of the 160 EUR median of the rest of the subscribers. However, the older subscribers tend to use less expensive phones. This tendency is also present within the football fans, but stronger within the other group.

\begin{figure}[ht]
    \centering
    \begin{subfigure}[t]{0.49\linewidth}
        \centering
        \includegraphics[width=\linewidth]{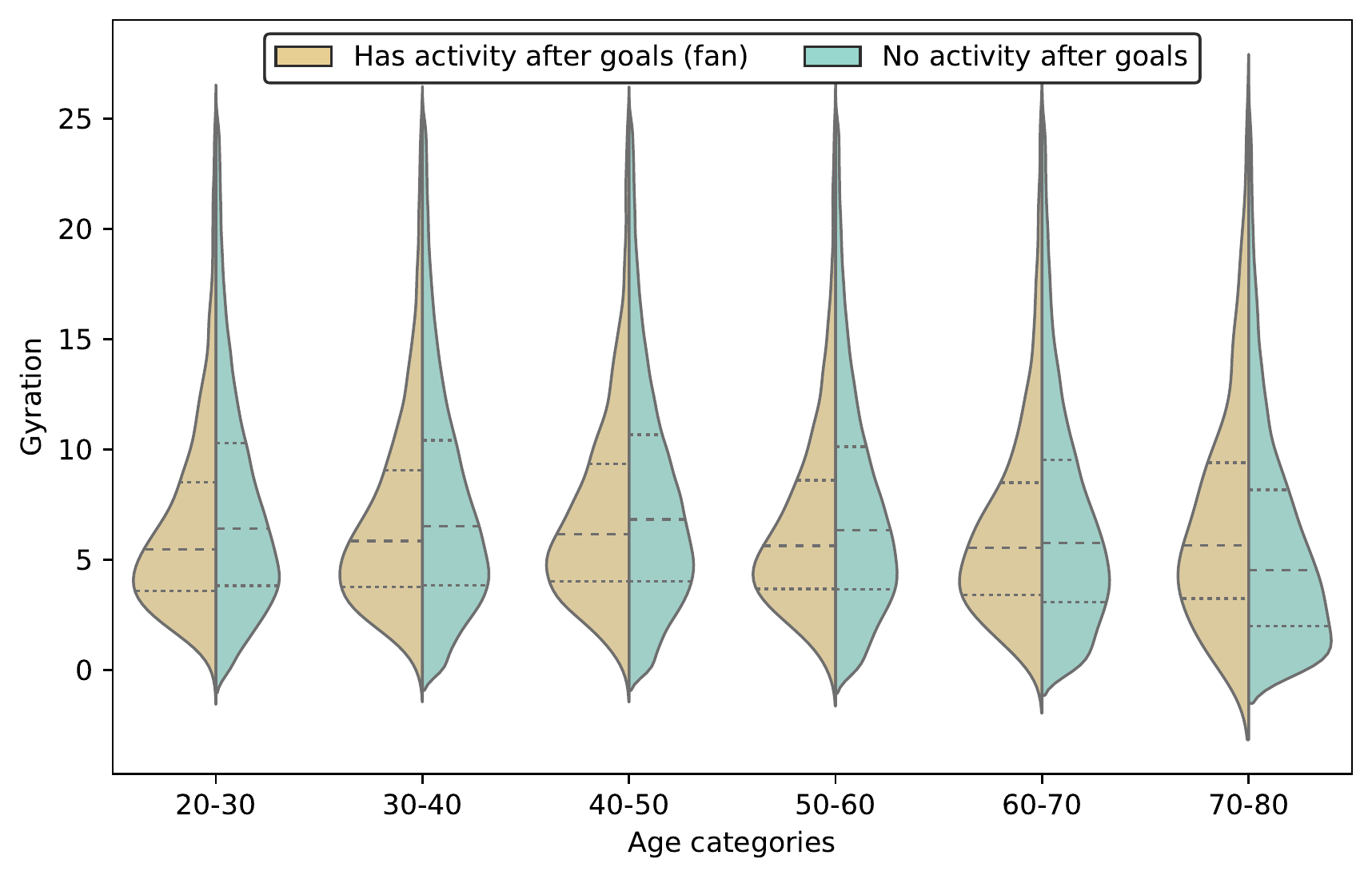}
        \caption{Radius of Gyration of the subscribers.}
        \label{fig:gyration_of_subscribers}
    \end{subfigure}
    \hfill
    \begin{subfigure}[t]{0.49\linewidth}
        \centering
        \includegraphics[width=\linewidth]{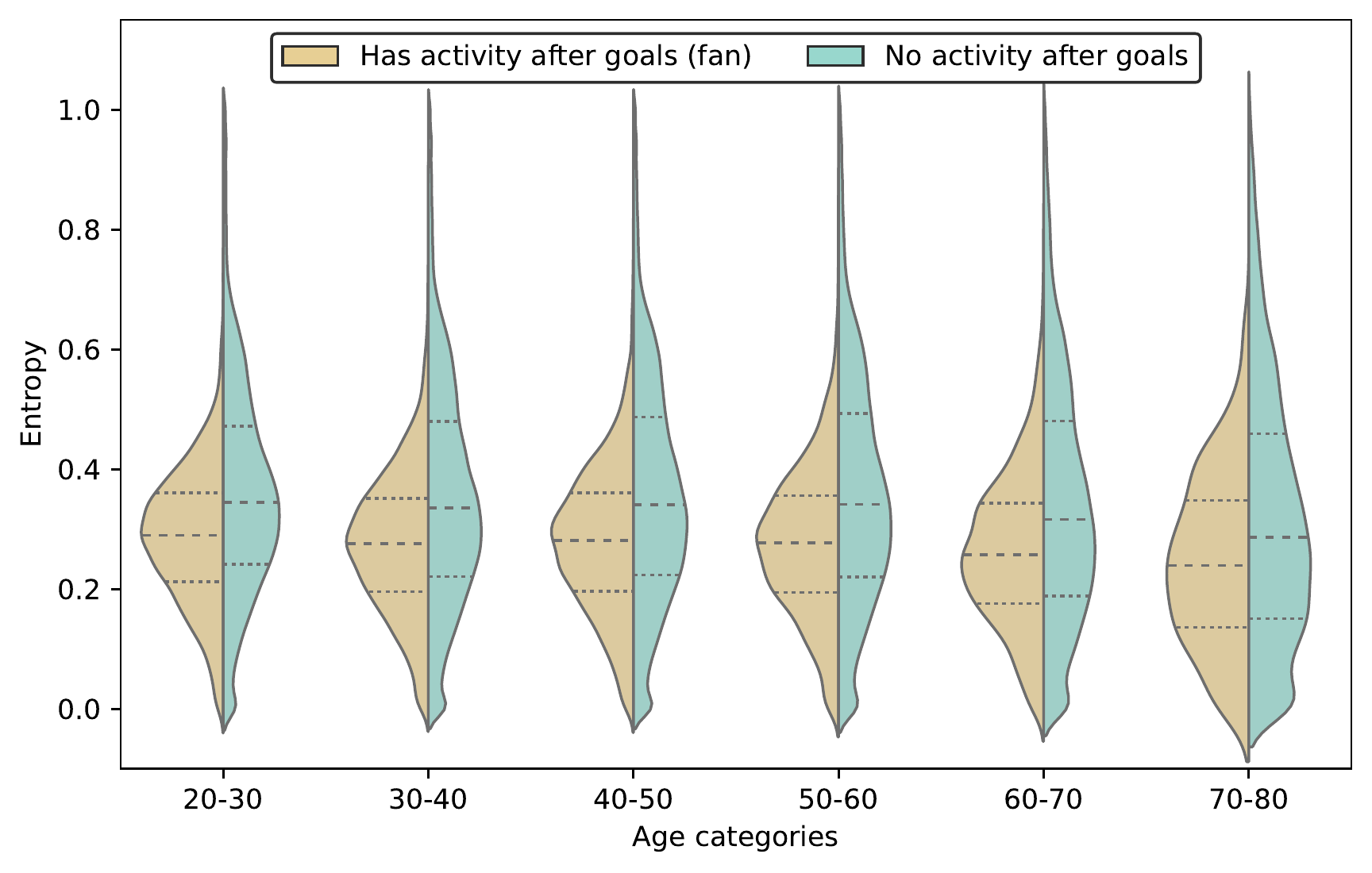}
        \caption{Entropy of the subscribers.}
        \label{fig:entropy_of_subscribers}
    \end{subfigure}

    \caption{Radius of Gyration and Entropy distributions in different age categories, comparing the fans, who had activity right after any two of the Hungarian goals, and the rest of the \acrshort{sim} cards.}
    \label{fig:gyration_and_entropy_of_subscribers}
\end{figure}

Figure~\ref{fig:gyration_and_entropy_of_subscribers}, illustrates the mobility metrics in different age categories, also comparing the football fans and the rest of the subscribers. The Radius of Gyration median is almost the same in all the age categories and groups. The Entropy medians have a notable difference between the two groups, but do not really change between the age categories.
This means, that the mobility customs of the football fans, who use the mobile phone network more actively, are similar, regardless of the subscribers' age.

\subsection{Hungary vs. Belgium}

\begin{figure}[ht]
    \centering
    \includegraphics[width=\linewidth]{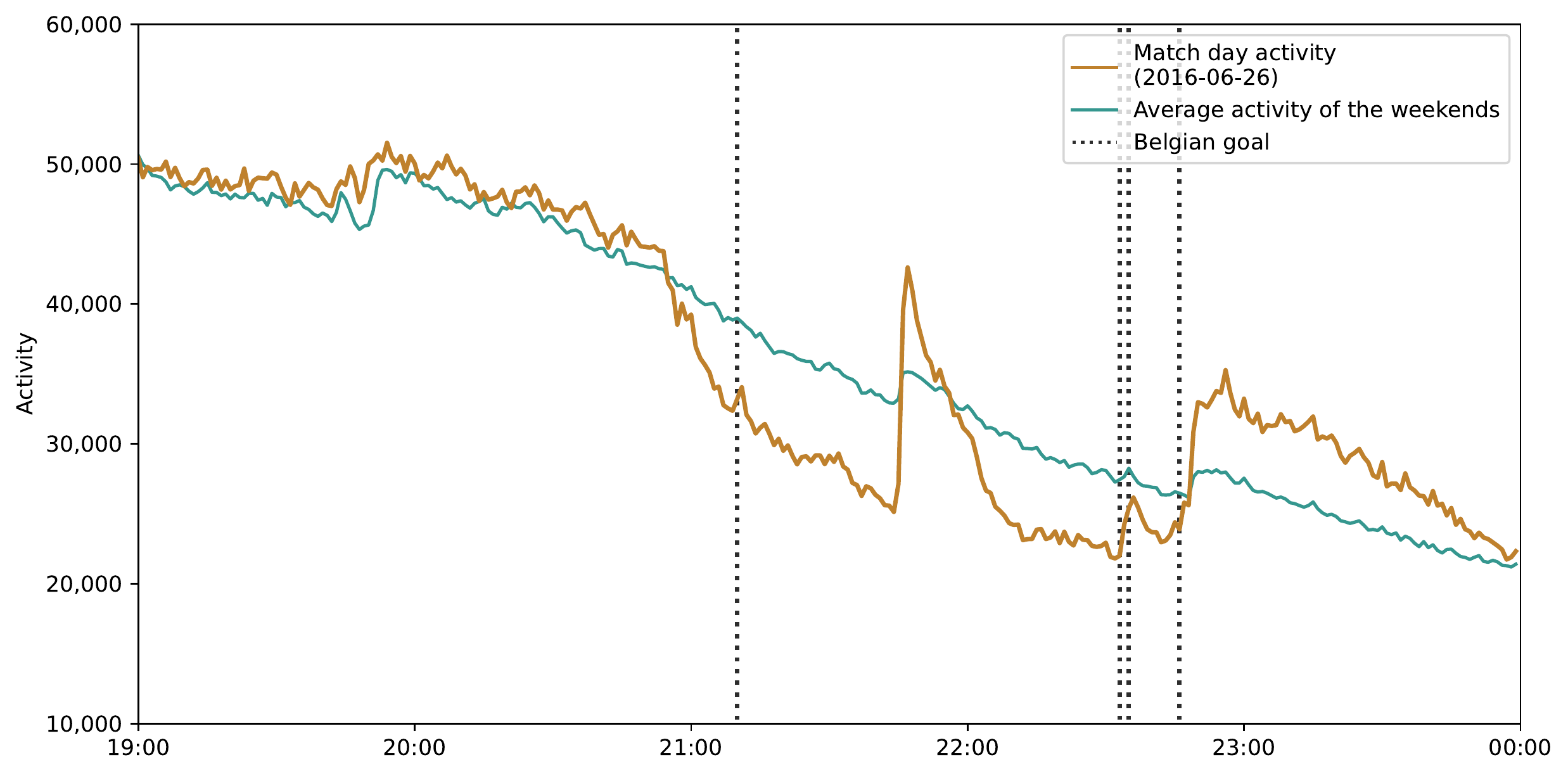}
    \caption{Mobile phone activity during and after the Hungary--Belgium Euro 2016 match, in comparison with the average activity of the weekends.}
    \label{fig:hun_bel_timeseries}
\end{figure}

On Sunday, June 26, 2016, Hungary played the fourth and last Euro 2016 match against Belgium. Figure~\ref{fig:hun_bel_timeseries}, shows the mobile phone network activity before, during and after the match.
During the match, the activity level was below the weekend average.
The activity after the match was slightly higher than average, since the match ended late on Sunday, when the activity average is usually very low. This activity surplus may only indicate that the fans were simply leaving the fan zones and going home.

\subsection{Homecoming}
\label{sec:homecoming}

The Hungarian national football team returned to Budapest, on June 27, 2016. A welcome event at the Heroes' Square have been held, where the football fans can greet the national football team. According to the M4 Sport television channel, approximately 20 thousand people attended to the event \cite{hiradohu2016tizezrek}.
Between 18:00 and 19:30, there were \num{4246} unique, non-phone \acrshort{sim} cards active in the site, that covers the Heroes' Square. \num{3425} are known to use smartphone, based on the operating system column of the GSMArena data set.

The cells of this base station cover a larger area, so not all of these subscribers actually attended to the event, but on the other hand, it is not compulsory to use the mobile phones during this event. Supposing that the mobile phone operator preferences among the attendees corresponded to the nationwide trends in 2016, there could even be about 17 thousand people, as the data provider had \num{25.3}\% market share \cite{nmhh_mobile_market_report}.

\begin{figure}[ht]
    \centering
    \begin{subfigure}[t]{0.5\linewidth}
        \centering
        \includegraphics[width=\linewidth]{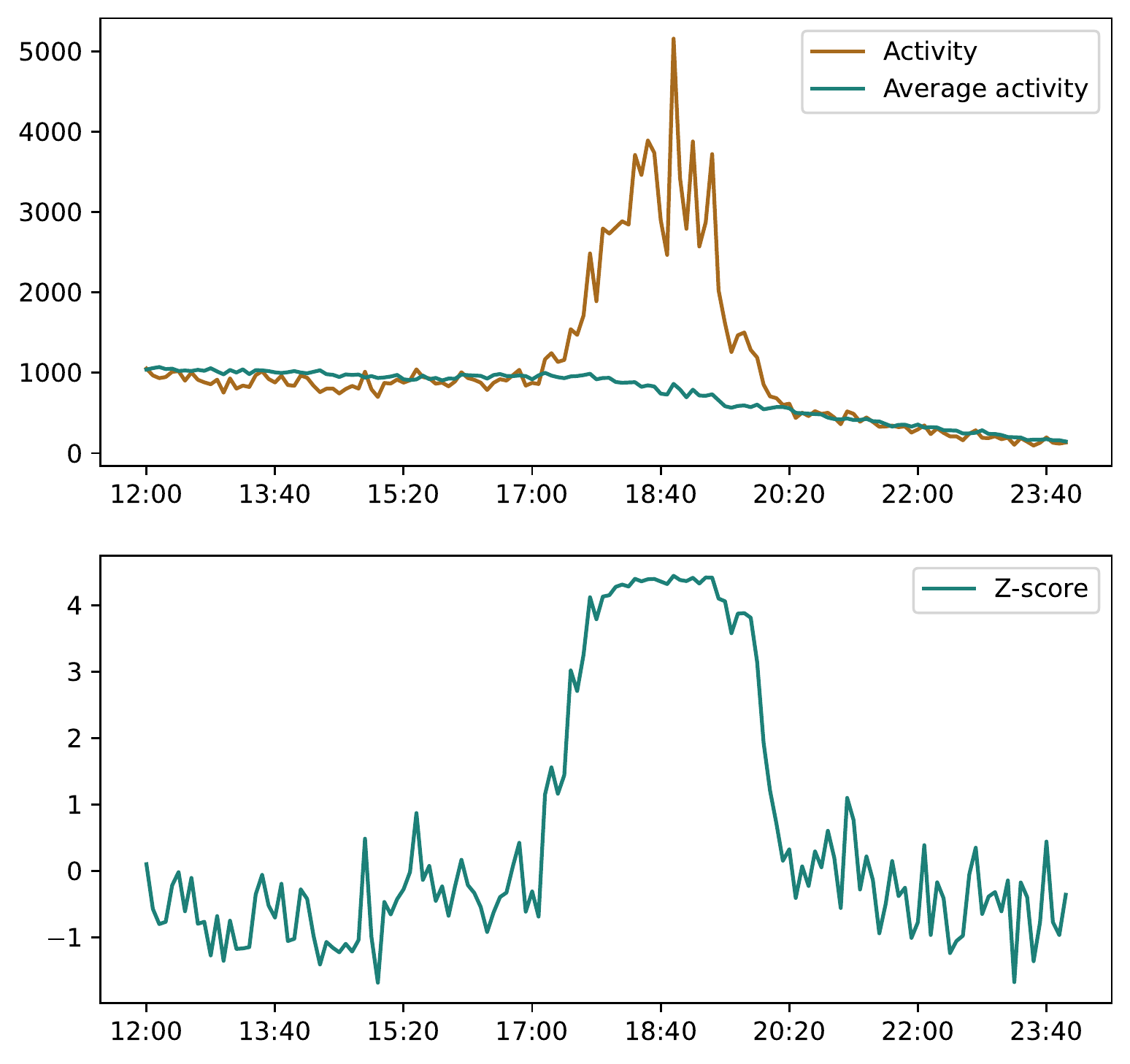}
        \caption{Activity and Z-score of the site, at Heroes' square.}
        \label{fig:heroes_square_welcoming_time_series}
    \end{subfigure}
    \hfill
    \begin{subfigure}[t]{0.475\linewidth}
        \centering
        \includegraphics[width=\linewidth]{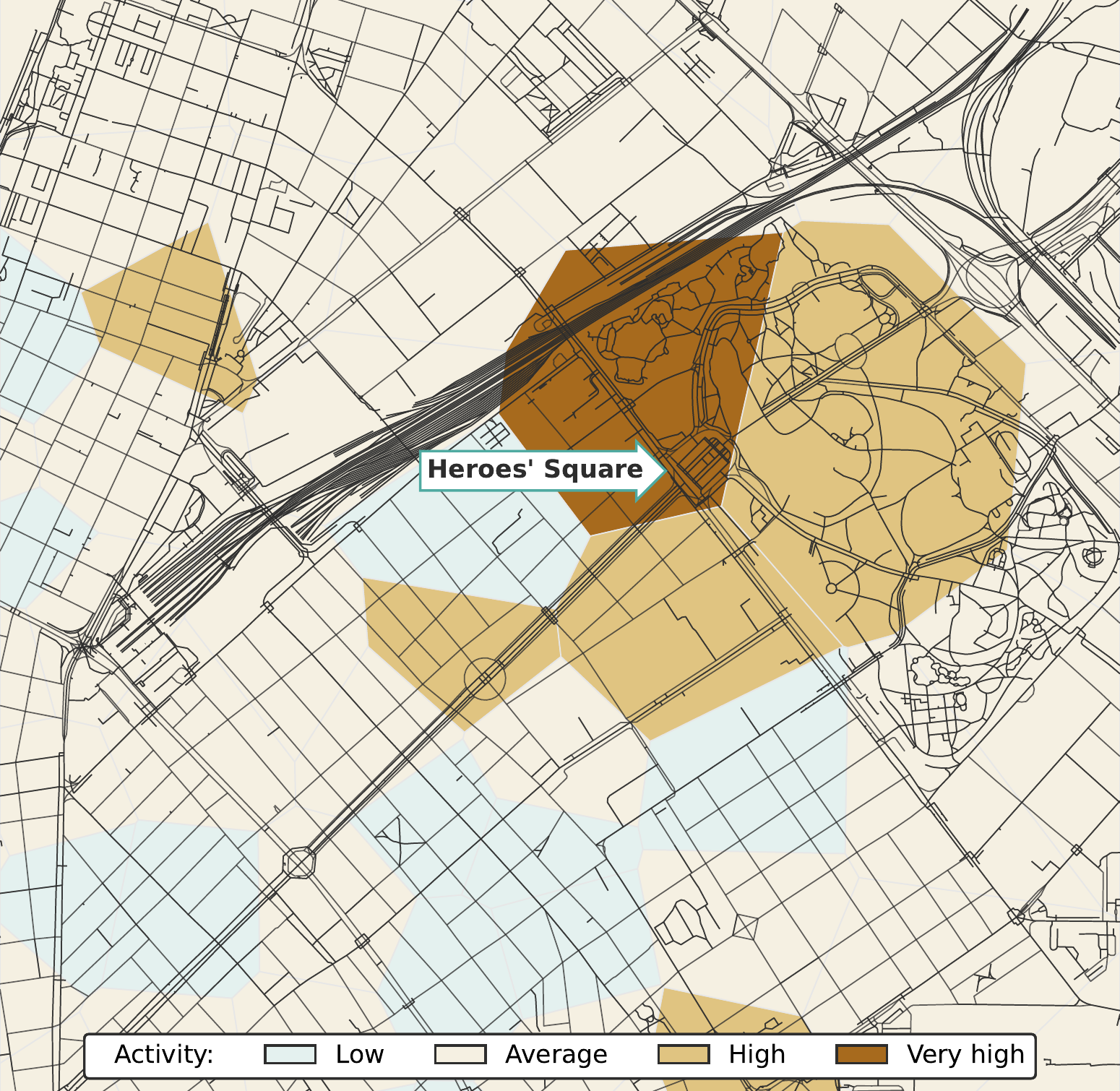}
        \caption{Spatial view of the activity at 18:35.}
        \label{fig:heroes_square_welcoming}
    \end{subfigure}

    \caption{Mobile phone network activity at Heroes' Square and its neighborhood, during the welcoming event of the Hungarian national football team.}
    \label{fig:welcoming}
\end{figure}

Figure~\ref{fig:heroes_square_welcoming} shows, a part of District 6 and the City Park with the Heroes' Square and the Voronoi polygons of the area are colored according to the Z-score values, to indicate the mobile phone activity in the area, at 18:35. The activity is considered low below $-1$, average between $-1$ and $1$, high between $1$ and $2.5$ and very high above $2.5$.
Figure~\ref{fig:heroes_square_welcoming_time_series} shows, the mobile phone network activity (upper), and the Z-score (bottom) of the site, covering Heroes' Square. It is clear, that during the event, the activity is significantly higher than the weekday average, and the Z-score values are also follows that.

\subsection{Limitations}

We associated subscribers' \acrshort{ses} with the release price of their cell phones, however, it is not necessary for them to buy their phones at that price. Many people buy their phone on sale or discount via the operator in exchange for signing an x-year contract.

Also, subscribers can change their phone devices at any time. We have taken into consideration only those subscribers, who had used only one device during the observation period, or had a dominant device that generated most of the activity records of the given subscriber.

We have fused three data sets to exclude the non-phone \acrshort{sim} cards, but the identified devices are not complete. There remained devices, that models are unknown and there are phones, that release date and price are unknown. It is not possible to determine \acrshort{ses} of these subscribers with the proposed solution.

\subsection{Future Work}

Although, the current solution to select the football fans' \acrshort{sim} cards, in other words, the \acrshort{sim} cards, that caused the peaks gives a reasonable result, but could be improved by analyzing the activity during the whole observation period. For example, applying a machine learning technique.

Extending the list of the non-phone \acrshort{tac}s could also help to refine the results, and combining the mobile phone prices with the real estate prices of the home location would most certainly enhance the socioeconomic characterization.

The relative age of the cell phone might be used as a weight for the phone price, when applied as \acrshort{ses} indicator to distinguish between the phone price categories. As an expensive, but older phone is not worth as much as a newer one with the same price.

\section{Conclusions}
\label{sec:conclusions}

In this study, we demonstrated that mobile phone network activity shadows precisely the football fans' behavior, even if the matches are played in another country. This analysis focused the people followed the matches on TV (at home) or big screens at the fan zones, but not in the stadium, where the matches were actually played.
The mobile phone network data and the mobile phone specification database has been applied to characterize the \acrshort{ses} of the football fans. The data fusion allowed us to remove a considerable number of \acrshort{sim} cards from the examination that certainly operates in other devices than mobile phones. Although, there are some still unidentified \acrshort{tac}s in the data set, but this way, the activity records, involved in this study, have a significantly higher possibility to used by an actual person during the events.

The time series of mobile network traffic clearly show that the activity was below the average during the matches, indicating that many people followed their team. This observation coincides with other studies \cite{traag2011social, mamei2016estimating, xavier2012analyzing, hiir2020impact}, where the activity of the cells at the stadium were analyzed.
We also demonstrated that a remote football match can also have notable effect on the mobile phone network.
Moreover, the joy felt after the Hungarian goals, is clearly manifested in the data, as sudden activity peaks.
The \acrshort{cdr} data is certainly capable of social sensing.

The spontaneous festival after the Hungary vs. Portugal match and the welcoming event at the Heroes' Square are direct applications of social sensing and comparable to mass protests from a data perspective. During the events, the mobile phone network activity was significantly higher than the average in affected areas.

The price of the mobile phone was proved to be an expressive socioeconomic indicator. It is capable not only to cluster the areas of a city, but also to distinguish the subscribers by mobility customs. On the other hand, it does not seem to affect the interest in football.

\vspace{6pt}

\section*{Author Contributions}
Conceptualization, G.P; methodology, G.P.; software, G.P.; validation, G.P.; formal analysis, G.P.; investigation, G.P.; resources, G.P. and I.F.; data curation, G.P.; writing---original draft preparation, G.P.; writing---review and editing, G.P. and I.F.; visualization, G.P.; supervision, I.F.; project administration, I.F.; funding acquisition, I.F. All authors have read and agreed to the published version of the manuscript.

\section*{Funding}
This research supported by the project 2019-1.3.1-KK-2019-00007 and by the Eötvös Loránd Research Network Secretariat under grant agreement no. ELKH KÖ-40/2020.

\section*{Acknowledgments}
The authors would like to thank Vodafone Hungary and 51Degrees for providing the Call Detail Records and the Type Allocation Code database for this study.

For plotting the map, OpenStreetMap data is used, that is copyrighted by the OpenStreetMap contributors and licensed under the Open Data Commons Open Database License (ODbL).

\section*{Conflicts of Interest}
The authors declare no conflict of interest. The funders had no role in the design of the study; in the collection, analyses, or interpretation of data; in the writing of the manuscript, or in the decision to publish the results.

\printglossary[title=Abbreviations, toctitle=Abbreviations, nogroupskip=true]

\printbibliography

\end{document}